\documentclass[12pt]{article}

\catcode`\@=11
\@addtoreset{equation}{section}

\global\arraycolsep=2pt

\usepackage{mathrsfs,amsbsy,amssymb,latexsym,amsfonts,amsmath,cite}
\usepackage{graphicx,color}

\newcommand{\al}{\alpha}
\newcommand{\be}{\beta}
\newcommand{\ga}{\gamma}

\newcommand{\ep}{\epsilon}

\newcommand{\Tr}{{\rm Tr}}

\newcommand{\el}{\nonumber}

\begin{document}

\begin{flushright}
\parbox{4cm}
{KUNS-2484 \\ 
ITP-UU-14/07 \\ 
SPIN-14/07}
\end{flushright}

\vspace*{1.5cm}

\begin{center}
{\Large\bf A Jordanian deformation of AdS space \\
in type IIB supergravity}
\vspace*{2.5cm}\\
{\large Io Kawaguchi$^{\ast}$\footnote{E-mail:~io@gauge.scphys.kyoto-u.ac.jp}, 
Takuya Matsumoto$^{\dagger}$\footnote{E-mail:~t.matsumoto@uu.nl} 
and Kentaroh Yoshida$^{\ast}$\footnote{E-mail:~kyoshida@gauge.scphys.kyoto-u.ac.jp}} 
\end{center}
\vspace*{0.25cm}
\begin{center}
$^{\ast}${\it Department of Physics, Kyoto University \\ 
Kyoto 606-8502, Japan.} 
\vspace*{0.25cm}\\ 
$^{\dagger}${\it Institute for Theoretical Physics and Spinoza Institute, 
Utrecht University, \\ 
Leuvenlaan 4, 3854 CE Utrecht, The Netherlands.} 
\end{center}
\vspace{1cm}

\begin{abstract}
We consider a Jordanian deformation of the AdS$_5\times$S$^5$ superstring 
action by taking a simple $R$-operator which satisfies the classical Yang-Baxter 
equation. The metric and NS-NS two-form are explicitly derived with a coordinate 
system. Only the AdS part is deformed and the resulting geometry 
contains the 3D Schr\"odinger spacetime as a subspace.  
Then we present the full solution in type IIB supergravity by determining 
the other field components. 
In particular, the dilaton is constant and a R-R three-form field strength is turned on. 
The symmetry of the solution is $\left[SL(2,\mathbb{R})\times U(1)^2\right]$ 
$\times$ $\left[SU(3)\times U(1)\right]$ and contains an anisotropic scale symmetry. 
\end{abstract}

\setcounter{footnote}{0}
\setcounter{page}{0}
\thispagestyle{empty}

\newpage

\section{Introduction}

One of the most intriguing subjects in string theory is the AdS/CFT correspondence \cite{M,GKP,W} 
and it has well been studied from various aspects with an enormous number of works. 
Although it is often supposed to hold as a matter of course in the recent studies,   
it is still important to elaborate the original form, 
the duality between type IIB string theory on AdS$_5\times$S$^5$ 
and the $\mathcal{N}$=4 super Yang-Mills (SYM) theory,  
to gain deeper insight for the basic origin of AdS/CFT. 
In this direction, the integrability behind this duality 
would play an important role (For a comprehensive review, see \cite{review}). 

\medskip 

We will concentrate on the string-theory side, type IIB string theory on AdS$_5\times$S$^5$ here. 
The Green-Schwarz type string action can be constructed based on a supercoset \cite{MT}, 
\[
PSU(2,2|4)/\left[SO(1,4) \times SO(5)\right]\,. 
\] 
This coset enjoys the $\mathbb{Z}_4$-grading property and it leads to the classical integrability 
\cite{BPR}\footnote{There is another formulation of 
the AdS$_5\times$S$^5$ superstring action \cite{RS}. 
For the classical integrability in this formalism, see \cite{Hatsuda}. }. 
The classification of possible supercosets, which 
lead to the classically integrable, consistent string theories, 
is performed in \cite{Zarembo-symmetric}\footnote{ 
Some new examples of AdS$_2$ and AdS$_3$ have been found in \cite{Wulf}.}. 

\medskip 

The next task is to consider integrable deformations. Although there are some kinds of integrable deformations, 
we will focus  upon $q$-deformations of the AdS$_5\times$S$^5$ superstring. 
Deformations of this type are the standard $q$-deformations of Drinfeld-Jimbo type 
\cite{Drinfeld1,Drinfeld2,Jimbo} (For a nice review, see \cite{CP}).  
The work along this direction was initiated by elaborating the classical integrability of 
squashed S$^3$ sigma models 
\cite{Cherednik,FR,BFP,Klimcik,Mohammedi,KY,KYhybrid,KMY-QAA,KMY-monodromy,ORU,KOY,BR}. 
Then the result was generalized to higher-dimensional cases \cite{DMV} 
with the help of the Yang-Baxter sigma model (YBsM) description \cite{Klimcik} 
(For recent progress on YBsM, see \cite{Squellari,bi-YB}). 

\medskip 

Then, by applying the YBsM description to the AdS$_5\times$S$^5$ superstring 
and the $q$-deformed classical action was presented in an abstract form 
with the group-theoretical language \cite{DMV2}. In the YBsM description, 
a linear $R$-operator is a key ingredient. It is constructed from a skew-symmetric, classical $r$-matrix 
satisfying the modified classical Yang-Baxter equation (mCYBE). The coordinate system has been introduced 
in \cite{ABF} and the metric in the string frame and NS-NS two-form have been determined. 
However, the complete gravitational solution has not been fixed yet in type IIB supergravity.

\medskip 

There is another kind of $q$-deformations, which are called Jordanian deformations 
\cite{R,Jordanian,KLM} or sometimes non-standard $q$-deformations 
(For the case of Lie superalgebras, see \cite{Tolstoy,BLT,ACS,ACCYZ}). 
In the previous work \cite{KMY-Jordanian-typeIIB}, 
we have considered Jordanian deformations 
of the AdS$_5\times$S$^5$ superstring action by using linear $R$-operators satisfying 
the classical Yang-Baxter equation (CYBE), rather than mCYBE. The action presented in \cite{KMY-Jordanian-typeIIB}
is also written abstractly in terms of a group element, and explicit examples have not been provided yet. 

\medskip 

In this paper, we consider a Jordanian deformation of the AdS$_5\times$S$^5$ with a simple $R$-operator. 
The metric and NS-NS two-form are explicitly derived with a coordinate 
system. Only the AdS part is deformed and the resulting geometry 
contains the 3D Schr\"odinger spacetime as a subspace.  
In this sense, this study can be regarded as a generalization of the previous works 
\cite{KY-Sch,Jordanian-KMY}. 
Then we present the full solution in type IIB supergravity by determining the other field components. 
In particular, the dilaton is constant and a R-R three-form field strength is turned on. 
The symmetry of the solution is given by $\left[SL(2,\mathbb{R})\times U(1)^2\right]$ 
$\times$ $\left[SU(3)\times U(1)\right]$ and contains an anisotropic scale symmetry.

\medskip 

This paper is organized as follows. 
In section 2 we give a short review of Jordanian deformations of 
the AdS$_5\times$S$^5$ superstring action. Then, by taking a simple $R$-operator satisfying CYBE, 
the metric and NS-NS two-form are explicitly derived with a coordinate system. 
The resulting geometry is given by the product of a deformed AdS space and round S$^5$\,. 
Also for a slightly generalized $R$-operator, the string action is derived. The resulting metric 
represents a time-dependent background. 
In section 3 we present the gravitational solution in type IIB supergravity by finding out  
the other field components. In particular, the dilaton is constant.  
Section 4 is devoted to conclusion and discussion. In Appendix A, our notation and convention 
is summarized. 
In Appendix B, we list some classical $r$-matrices and the associated string actions.

\section{Jordanian deformations of AdS$_5\times$S$^5$} 

In this section, we first introduce Jordanian deformations of the AdS$_5\times$S$^5$ superstring. 
Then by taking a simple example of skew-symmetric, classical $r$-matrix, the string action is obtained 
with a coordinate system. Then the metric and NS-NS two-form are derived explicitly. The resulting metric contains 
the 3D Schr\"odinger spacetime as a subspace. 
A more general example is also presented. 

\subsection{Setup}

First of all, we will give a short summary of the work \cite{KMY-Jordanian-typeIIB}. 
One may consider Jordanian deformations of the AdS$_5\times$S$^5$ superstring action with 
linear $R$ operators satisfying CYBE. The construction follows basically \cite{DMV2} with the help of 
the YBsM description \cite{Klimcik}.  

\medskip 

The deformed Green-Schwarz string action is given by \cite{KMY-Jordanian-typeIIB}
\begin{eqnarray}
S=-\frac{1}{2}\int^\infty_{-\infty}\!\!\!d\tau\int^{2\pi}_0\!\!\!d\sigma~
P_-^{\alpha\beta}{\rm Str}\left(A_\alpha d\circ\frac{1}{1-\eta\left[R_{\rm Jor}\right]_g\circ d}(A_\beta)\right)\,,  
\label{action}
\end{eqnarray}
where the left-invariant one-form $A_\alpha$ is given by 
\begin{eqnarray}
A_\alpha\equiv g^{-1}\partial_\alpha g\,, \qquad 
g\in SU(2,2|4)\,. 
\end{eqnarray}
The projection operators $P_\pm^{\alpha\beta}$ are defined as linear-combinations of 
the metric $\gamma^{\alpha\beta}$ and the anti-symmetric tensor $\epsilon^{\alpha\beta}$ 
on the string world-sheet like  
\begin{eqnarray}
P_\pm^{\alpha\beta}\equiv\frac{1}{2}\left(\gamma^{\alpha\beta}\pm\epsilon^{\alpha\beta}\right)  
\end{eqnarray}
and satisfy the following properties,  
\begin{eqnarray}
P_\pm^{\alpha\gamma}\gamma_{\gamma\delta}P_{\pm}^{\delta\beta}=P_\pm^{\alpha\beta}\,, \qquad
P_\pm^{\alpha\gamma}\gamma_{\gamma\delta}P_{\mp}^{\delta\beta}=0\,. \qquad
\end{eqnarray}
In the action (\ref{action}) the projection $P_{-}^{\alpha\beta}$ is utilized. 

\medskip 

Recall that the Lie superalgebra $\mathfrak{su}(2,2|4)$ has a ${\mathbb Z}_4$ automorphism, 
$\Omega$ with $\Omega^4=1$\,. 
This automorphism leads to the decomposition of $\mathfrak{su}(2,2|4)$ as follows:  
\begin{eqnarray}
\mathfrak{su}(2,2|4)&=&\mathfrak{su}(2,2|4)^{(0)}\oplus \mathfrak{su}(2,2|4)^{(1)}\oplus 
\mathfrak{su}(2,2|4)^{(2)}\oplus \mathfrak{su}(2,2|4)^{(3)}\,.
\end{eqnarray}

Here the operation of $\Omega$ is defined for an element of $\mathfrak{su}(2,2|4)^{(n)}$ as 
\begin{eqnarray}
&&\Omega(X^{(n)})=i^nX^{(n)} \qquad \mbox{for}~~X^{(n)}\in \mathfrak{su}(2,2|4)^{(n)}\,.
\end{eqnarray}
and $\mathfrak{su}(2,2|4)^{(n)}$ satisfies the following relation, 
\begin{eqnarray}
&&\left[\mathfrak{su}(2,2|4)^{(n)},\,\mathfrak{su}(2,2|4)^{(m)}\right]\subset \mathfrak{su}(2,2|4)^{(n+m)} 
\quad (\mbox{mod}~4)\,. \nonumber
\end{eqnarray}
In particular, $\mathfrak{su}(2,2|4)^{(0)}$ is nothing but $\mathfrak{so}(1,4)\times \mathfrak{so}(5)$\,. 

\medskip 

One can introduce projections $P_n$ from $\mathfrak{su}(2,2|4)$ onto $\mathfrak{su}(2,2|4)^{(n)}$ ($n=0,1,2,3$)\,. 
Then the operator $d$ is defined as 
\begin{eqnarray}
d \equiv P_1+2P_2-P_3\,. 
\end{eqnarray}

\medskip 

The remaining task is to introduce the operation $\left[R_{\rm Jor}\right]_g$\,. 
In the first place, we introduce a linear $R$-operator, $R_{\rm Jor}$\,,  
\begin{eqnarray}
R_{\rm Jor}~:~~\mathfrak{gl}(4|4) \to \mathfrak{gl}(4|4)\,, 
\end{eqnarray}
which satisfies the following three properties, 
\begin{enumerate}
\item[1)] \quad the classical Yang-Baxter equation (CYBE)\,: 
\begin{eqnarray}
\bigl[R_{\rm Jor}(M),R_{\rm Jor}(N)\bigr]
-R_{\rm Jor}\left([R_{\rm Jor}(M),N]+[M,R_{\rm Jor}(N)]\right)=0\,, \nonumber 
\end{eqnarray}
\item[2)] \quad the nilpotency\,: \qquad~~~ $(R_{\rm Jor})^n(M)=0 \qquad (n \geq 3)$\,, 
\item[3)] \quad the skew-symmetric property (the unitarity condition): 
\begin{eqnarray}
{\rm Str}(MR_{\rm Jor}(N))=-{\rm Str}(R_{\rm Jor}(M)N)\,. \nonumber 
\end{eqnarray}
\end{enumerate} 
The property 3) is not necessary for the definition of Jordanian deformations, 
but for the classical integrability of the string theory.

\medskip 

Note that $R_{\rm Jor}$  does not preserve the real-form condition 
of $\mathfrak{su}(2,2|4)$ in general, even if the domain is restricted 
to $\mathfrak{su}(2,2|4)$\,. The real-form condition is not necessary 
for the classical integrability (i.e., the construction of Lax pair) 
as shown in \cite{KMY-Jordanian-typeIIB}. One may expect that the string actions become complex 
if the real-form condition is not preserved. However, it is not always the case. 
In fact, some examples of $R_{\rm Jor}$\,, which break the real-form condition, 
give rise to real string actions, as we will see later. 
That is, the real-form condition should be regarded as a sufficient condition for the reality.  
Still, we have no general criterion to specify the linear operators that lead to the real string actions. 
It is an important issue to argue the criterion in the future. 

\medskip 

Then the operator $\left[R_{\rm Jor}\right]_g$ is defined as a sequence 
of the adjoint operation Ad$_g$ by $g$\,, the $R$-operation and the inverse of the adjoint\,: 
\begin{eqnarray}
\left[R_{\rm Jor}\right]_g(M)\equiv {\rm Ad}_g^{-1}\circ R_{\rm Jor}\circ {\rm Ad}_g(M)=g^{-1}R_{\rm Jor}(gMg^{-1})g\,. 
\end{eqnarray}
This operation is intrinsic to the coset case \cite{Klimcik,DMV}. 

\subsubsection*{The tensorial notation of the $R$-operator}

It is helpful to see the tensorial notation of $R_{\rm Jor}$\,. In the present case, 
$r_{\rm Jor}$ is skew-symmetric 
due to the property 3)\,. Hence $r_{\rm Jor}$ can be represented by using a skew-symmetrized 
tensor product of two elements of $\mathfrak{gl}(4|4)$\,, 
\begin{eqnarray}
r_{\rm Jor}=\sum_i(a_i\otimes b_i-b_i\otimes a_i)\equiv\sum_ia_i\wedge b_i\,. 
\end{eqnarray} 
The linear $R$ operator action is associated with the tensorial notation as follows:  
\begin{eqnarray}
R_{\rm Jor}(M)\equiv{\rm Tr}_2\left[r_{\rm Jor}(1\otimes M)\right]
=\sum_i \left(a_i {\rm Tr}(b_iM)-b_i {\rm Tr}(a_iM) \right)\,. 
\label{relation of r and R}
\end{eqnarray}

In the tensorial notation, the property 1) is recast into the familiar expression of CYBE,  
\begin{eqnarray}
\left[r_{{\rm Jor},12},r_{{\rm Jor},13}\right]
+\left[r_{{\rm Jor},12},r_{{\rm Jor},23}\right]+
\left[r_{{\rm Jor},13},r_{{\rm Jor},23}\right]=0\,. 
\end{eqnarray}
Here the subscripts of $r_{\rm Jor}$ specify vector spaces on which $r_{\rm Jor}$ acts.  

\medskip 

Thus a skew-symmetric solution of CYBE is associated with a linear $R$-operator 
and is related to an integrable deformation of AdS$_5\times$S$^5$\,. 
So far, the string action (\ref{action}) is written in an abstract form with a group-theoretical language. 
In the next subsection, we will take an example of skew-symmetric classical $r$-matrix 
and express explicitly the action with a coordinate system. 

\subsection{A simple example of the string action}

Let us consider an explicit example of Jordanian deformations 
by taking a skew-symmetric classical $r$-matrix\footnote{The term ``Jordanian'' comes from 
the fact that $r_{\rm Jor}$ is represented by an upper triangular matrix.},   
\begin{eqnarray}
r_{\rm Jor} 
&=&\frac{1}{\sqrt{2}}E_{24}\wedge(E_{22}-E_{44})\,,  \label{r-1}
\end{eqnarray}
where $E_{ij}$ is a $4\times 4$ matrix defined as 
\[
(E_{ij})_{kl} \equiv \delta_{ik}\delta_{jl}\,.
\] 
The normalization of $r_{\rm Jor}$ is absorbed by rescaling 
of $\eta$\,, as one can see from the action (\ref{action})\,. 
Here it is fixed for later convenience. 
 
 \medskip 
 
The $r$-matrix (\ref{r-1}) induces the action of the associated $R$-operator as 
\begin{eqnarray}
R_{\rm Jor}(E_{22})=-R_{\rm Jor}(E_{44})=\frac{1}{\sqrt{2}}E_{24}\,, \qquad 
R_{\rm Jor}(E_{42})=-\frac{1}{\sqrt{2}}(E_{22}-E_{44})\,. 
\end{eqnarray}
This mapping rule is obtained from the relation \eqref{relation of r and R}\,. 
Note that the $r$-matrix (\ref{r-1}) does not preserve the real-form condition of $\mathfrak{su}(2,2|4)$\,. 
However, it leads to a real string action, as we will see later. 

\medskip 

Let us evaluate the string action (\ref{action})\,. For simplicity, we focus on the bosonic part 
by restricting a group element $g$ to the bosonic subsector. 
In addition, only the AdS$_5$ part is deformed in the present example and 
hence the coset construction for the S$^5$ part is the usual. 
Therefore, we concentrate on the coset 
construction for the AdS$_5$ part. 
Then it is convenient to consider the following coset representative,
\begin{eqnarray}
g ={\rm e}^{p_0x^0+p_1x^1+p_2x^2+p_3x^3}{\rm e}^{\gamma_5\rho/2} ~~\in~ SU(2,2)/SO(1,4)\,. 
\label{Poincare coordinate}
\end{eqnarray}
Here $p_\mu$\,~($\mu=0,1,2,3$) are defined as 
\begin{eqnarray}
&& p_0=\frac{1}{2}\left(\gamma_0-2n_{05}\right)\,, \qquad 
p_1=\frac{1}{2}\left(\gamma_1-2n_{15}\right)\,, \nonumber \\ 
&& p_2=\frac{1}{2}\left(\gamma_2-2n_{25}\right)\,, \qquad 
p_3=\frac{1}{2}\left(\gamma_3-2n_{35}\right)\,. 
\end{eqnarray}
For the definition of the $\mathfrak{so}(2,4)$ generators $\gamma_\mu$\,, $\gamma_5$\,, 
$n_{\mu\nu}$ and $n_{\mu 5}$\,,  
see Appendix A. 

\medskip 

Later, we often use the following quantities 
\begin{eqnarray}
x^\pm\equiv\frac{1}{\sqrt{2}}\left(x^0\pm x^3\right)\,, \qquad
z\equiv{\rm e}^\rho\,, 
\end{eqnarray}
instead of $x^0$\,, $x^3$ and $\rho$\,. 

\medskip 

With this setup, the classical action (\ref{action}) can be rewritten as 
\begin{eqnarray}
S&=&S_{\rm AdS}+S_{\rm S}\,, \nonumber \\
S_{\rm AdS}&=&-\frac{1}{2}\int^\infty_{-\infty}\!\!\!d\tau\int^{2\pi}_0\!\!\!d\sigma~
(\gamma^{\alpha\beta}-\epsilon^{\alpha\beta}){\rm Tr}\left(A_\alpha P_2\circ
\frac{1}{1-2\eta\left[R_{\rm Jor}\right]_g\circ P_2}(A_\beta)\right)\,. 
\end{eqnarray}
Here $A_\alpha$ is restricted to $\mathfrak{su}(2,2)$\,. Then $S_{\rm S}$ represents the usual S$^5$ part of 
the string action and we will not touch on this in the present section. 

\medskip 

From now on, let us compute the explicit form of $A_\alpha$ and 
\begin{eqnarray}
J_\alpha \equiv \frac{1}{1-2\eta\left[R_{\rm Jor}\right]_g\circ P_2}(A_\alpha)\,. 
\label{definitionJ}
\end{eqnarray}
After that, the bosonic part of the classical action can be determined explicitly with the coordinate system 
introduced with the parametrization (\ref{Poincare coordinate})\,. 

\medskip 

First of all, 
$P_2(A_\alpha)$ can be evaluated as 
\begin{eqnarray}
&&P_2(A_\alpha)=\gamma_0\,a_\alpha^0 + \gamma_1\,a_\alpha^1+\gamma_2\,a_\alpha^2
+\gamma_3\,a_\alpha^3+\gamma_5\,a_\alpha^5\,, 
\label{P_2(A)} 
\end{eqnarray}
where each of the coefficients is given by 
\begin{eqnarray}
&&a_\alpha^1=\frac{\partial_\alpha x^1}{2z}\,, \qquad
a_\alpha^2=\frac{\partial_\alpha x^2}{2z}\,, \qquad 
\frac{1}{\sqrt{2}}(a_\alpha^3+a_\alpha^0)=\frac{\partial_\alpha x^+}{2z}\,, \nonumber \\ 
&& \frac{1}{\sqrt{2}}(a_\alpha^3-a_\alpha^0)=-\frac{\partial_\alpha x^-}{2z}\,, \qquad 
a_\alpha^5=\frac{\partial_\alpha z}{2z}\,. \nonumber
\end{eqnarray}
The next task is to evaluate $P_2(J_\alpha)$\,. 
The relation \eqref{definitionJ} can be inverted as 
\begin{eqnarray}
A_\alpha=\left(1-2\eta\left[R_{\rm Jor}\right]_g\circ P_2\right)(J_\alpha)\,. 
\end{eqnarray}
By acting $P_2$ on both sides, the following expression is obtained, 
\begin{eqnarray}
P_2(A_\alpha)&=&P_2\circ\left(1-2\eta\left[R_{\rm Jor}\right]_g\circ P_2\right)(J_\alpha) \nonumber \\
&=&P_2(J_\alpha)-2\eta P_2\circ\left[R_{\rm Jor}\right]_g(P_2(J_\alpha))\,. \label{linear equation}
\end{eqnarray}
This is just a linear equation for $P_2(J_\alpha)$ and hence it is straightforward to evaluate 
$P_2(J_\alpha)$\,. 
Note that $P_2(J_\alpha)$ can be expanded as 
\begin{eqnarray}
&&P_2(J_\alpha)=\gamma_0\,j_\alpha^0+\gamma_1\,j_\alpha^1+\gamma_2\,j_\alpha^2
+\gamma_3\,j_\alpha^3+\gamma_5\,j_\alpha^5\,. 
\end{eqnarray}
Here $j_{\alpha}^{\mu}$ and $j_{\alpha}^5$ are unknown functions to be determined. 
With this expansion, the right-hand side of \eqref{linear equation} can be rewritten as 
\begin{eqnarray}
P_2(A_\alpha)&=&\gamma_1\frac{2z^2j_\alpha^1+\sqrt{2}\eta x^1(j_\alpha^3+j_\alpha^0)}{2z^2}
+\gamma_2\frac{2z^2j_\alpha^2+\sqrt{2}\eta x^2(j_\alpha^3+j_\alpha^0)}{2z^2} \nonumber \\
&&+\gamma_3\frac{2z^2j_\alpha^3-\sqrt{2}\eta(x^1j_\alpha^1+x^2j_\alpha^2+zj_\alpha^5)}{2z^2}
+\gamma_0\frac{2z^2j_\alpha^0+\sqrt{2}\eta(x^1j_\alpha^1+x^2j_\alpha^2+zj_\alpha^5)}{2z^2} \nonumber \\
&&+\gamma_5\frac{2zj_\alpha^5+\sqrt{2}\eta(j_\alpha^3+j_\alpha^0)}{2z}\,. \label{this}
\end{eqnarray}
By comparing (\ref{this}) with \eqref{P_2(A)}\,, 
the expressions of $j_{\alpha}^{\mu}$ and $j_{\alpha}^5$ are determined as 
\begin{eqnarray}
&&j_\alpha^1=\frac{z^2\partial_\alpha x^1-\eta x^1\partial_\alpha x^+}{2z^3}\,, \qquad
j_\alpha^2=\frac{z^2\partial_\alpha x^2-\eta x^2\partial_\alpha x^+}{2z^3}\,, \nonumber \\
&&\frac{1}{\sqrt{2}}(j_\alpha^3-j_\alpha^0)=\frac{-z^2\partial_\alpha x^-
-\eta(x^1\partial_\alpha x^1+x^2\partial_\alpha x^2+z\partial_\alpha z)
-\eta^2\left(1+\frac{(x^1)^2+(x^2)^2}{z^2}\right)\partial_\alpha x^+}{2z^3}\,, \nonumber \\
&&\frac{1}{\sqrt{2}}(j_\alpha^3+j_\alpha^0)=\frac{\partial_\alpha x^+}{2z}\,, \qquad 
j_\alpha^5=\frac{z\partial_\alpha z-\eta \partial_\alpha x^+}{2z^2}\,. \nonumber
\end{eqnarray}
Thus the classical action has been obtained as 
\begin{eqnarray}
S_{\rm AdS}&=&-\frac{1}{2}\int^\infty_{-\infty}\!\!\!d\tau\!\int^{2\pi}_0\!\!\!d\sigma\,
\gamma^{\alpha\beta}
\Biggl[\frac{1}{z^2}\left(-2\partial_\alpha x^+\partial_\beta x^-
+\partial_\alpha x^1\partial_\beta x^1
+\partial_\alpha x^2\partial_\beta x^2 +\partial_\alpha z\partial_\beta z\right) \Biggr. \nonumber \\
&&\hspace{4.5cm}\Biggl.
-\frac{\eta^2}{z^4}\left(1+\frac{(x^1)^2+(x^2)^2}{z^2}\right)\partial_\alpha x^+\partial_\beta x^+\Biggr] \nonumber \\
&&+\int^\infty_{-\infty}\!\!\!d\tau\!\int^{2\pi}_0\!\!\!d\sigma\,
\epsilon^{\alpha\beta}\frac{\eta}{z^4}
\Bigl(x^1\partial_\alpha x^+\partial_\beta x^1
+x^2\partial_\alpha x^+\partial_\beta x^2
+z\partial_\alpha x^+\partial_\beta z\Bigr)\,, \label{action-1}
\end{eqnarray}
where the last term in the coupling to NS-NS two-form is a surface term 
and it can be dropped off. 
This action is real. 
From this action and the S$^5$ part, one can read off the metric in the string frame 
and NS-NS two-form. In order to determine the string background completely, it is still 
necessary to fix the other field components by solving the field equations of motion 
in type IIB supergravity. This will be the issue in the next section. 

\medskip 

So far, the $r$-matrix (\ref{r-1}) has been considered. 
Note that the following four $r$-matirces 
\begin{eqnarray}
&& r_{\rm Jor}^{(1)} = \frac{1}{\sqrt{2}}E_{13} \wedge (E_{11}-E_{33})\,, \quad 
r_{\rm Jor}^{(2)} = \frac{1}{\sqrt{2}}E_{23} \wedge (E_{22}-E_{33})\,, \nonumber \\ 
&& r_{\rm Jor}^{(3)} = \frac{1}{\sqrt{2}} E_{14} \wedge (E_{11}-E_{44})\,,  \quad 
r_{\rm Jor}^{(4)} = \frac{1}{\sqrt{2}}\bigl[E_{14} \wedge (E_{11}-E_{44}) - 2 (E_{12}\wedge E_{24} + E_{13} \wedge E_{34})\bigr] \nonumber 
\end{eqnarray}
lead to the same string action (\ref{action-1})\,, up to double Wick rotations and coordinate transformations 
(For the detail, see Appendix B). 
Note that $r_{\rm Jor}^{(2)}$ and $r_{\rm Jor}^{(4)}$ are obtained by performing 
adjoint operations with $\Delta(E_{23})$ and $\Delta(E_{14})$, respectively,   
to the classical $r$-matrix of Drinfeld-Jimbo type satisfying mCYBE,  
as argued in \cite{KMY-Jordanian-typeIIB}. Thus the present example 
may be regarded as Jordanian twists \cite{R,Jordanian,KLM}, 
though this fact is not manifest from the expression of the $r$-matrix (\ref{r-1})\,.

\subsection{Other examples}

Before closing this section, let us present a generalized example\footnote{The imaginary unit $i$ is multiplied 
so that the resulting NS-NS two-form should be real. } 
of skew-symmetric $r$-matrix satisfying CYBE, 
\begin{eqnarray}
r_{\rm Jor}=\frac{i}{\sqrt{2}} \Bigl[ 
E_{24}\wedge(E_{22}-E_{44})-2E_{23}\wedge E_{34}
\Bigr]\,. \label{r-2}
\end{eqnarray}
This $r$-matrix is 
also regarded as a Jordanian twist \cite{R,Jordanian,KLM}.  
Note that the $r$-matrix (\ref{r-2}) does not preserve the real-form condition of $\mathfrak{su}(2,2|4)$\,. 
However, this is also an example which gives rise to a real string action, as we will see below. 

\medskip 

We will not show the derivation in detail. With the $r$-matrix (\ref{r-2}), only the AdS$_5$ part is deformed 
again and the algorithm of the derivation is the same. 

\medskip 

The resulting action for the deformed AdS$_5$ part is given by 
\begin{eqnarray}
S_{\rm AdS}&=&-\frac{1}{2}\int^\infty_{-\infty}\!\!\!d\tau\int^{2\pi}_0\!\!\!d\sigma~
\gamma^{\alpha\beta}\frac{z^2}{z^4+4\eta^2(x^+)^2} \nonumber \\
&&\qquad\times\Bigl[-2\partial_\alpha x^+\partial_\beta x^-
+\partial_\alpha x^1\partial_\beta x^1
+\partial_\alpha x^2\partial_\beta x^2
+\partial_\alpha z\partial_\beta z 
\Bigr. \nonumber \\
&&\hspace{1.5cm}
+\frac{\eta^2}{z^4}\bigl(
4x^+\partial_\alpha x^+\left(x^1\partial_\beta x^1+x^2\partial_\beta x^2-2x^+\partial_\beta x^-\right)
+4(x^+)^2\partial_\alpha z\partial_\beta z\bigr. \nonumber \\
&&\hspace{1.5cm}\bigl.+\left(z^2-(x^1)^2-(x^2)^2\right)\partial_\alpha x^+\partial_\beta x^+
\bigr)
+4\frac{\eta^4}{z^6}(x^+)^2\partial_\alpha x^+\partial_\beta x^+\Bigr] \nonumber \\
&&+\int^\infty_{-\infty}\!\!\!d\tau\int^{2\pi}_0\!\!\!d\sigma~
\epsilon^{\alpha\beta}\frac{\eta}{z^4+4\eta^2(x^+)^2} \nonumber \\
&&\qquad\times\Bigl[
x^2\partial_\alpha x^+\partial_\beta x^1
-x^1\partial_\alpha x^+\partial_\beta x^2
+2x^+\partial_\alpha x^1\partial_\beta x^2\Bigr]\,, \label{action-2}
\end{eqnarray}
where surface terms have already been ignored. This action is real again. 
Note that the NS-NS two-form contains an imaginary part if the surface term 
cannot be dropped off, for example, by considering the open string case.  
In the present case, the metric depends on the light-cone time $x^+$ explicitly. 
This metric may have an interesting feature as a dynamical brane background.  

\subsubsection*{A two-parameter deformation}

It is also interesting to see a two-parameter deformation of AdS$_5$\,.  
It can be considered by taking the following $r$-matrix, 
\begin{eqnarray}
r_{\rm Jor}=\frac{s_1}{\sqrt{2}}E_{24}\wedge(E_{22}-E_{44})
+\frac{s_2}{\sqrt{2}}E_{13}\wedge(E_{11}-E_{33})\,,
\end{eqnarray}
where $s_1$ and $s_2$ are constant parameters. The resulting string action is given by 
\begin{eqnarray} 
S_{\rm AdS} &=&-\frac{1}{2}\int^\infty_{-\infty}\!\!\!d\tau\!\int^{2\pi}_0\!\!\!d\sigma\,
\gamma^{\alpha\beta}\frac{z^2}{z^4+2\eta^2s_1s_2 \left[z^2+(x^1)^2+(x^2)^2\right]} \nonumber \\
&& \times\Bigl(-2\partial_\alpha x^+\partial_\beta x^-
+\partial_\alpha x^1\partial_\beta x^1
+\partial_\alpha x^2\partial_\beta x^2
+\partial_\alpha z\partial_\beta z \Bigr. \nonumber \\
&&\hspace{0.5cm}\Bigl.-\frac{\eta^2}{z^2}\Bigl[
\bigl(1+\frac{(x^1)^2+(x^2)^2}{z^2}\bigr)
\left(s_1\partial_\alpha x^+ + s_2\partial_\alpha x^-\right)
\left(s_1\partial_\beta x^+ + s_2\partial_\beta x^-\right)\nonumber \\
&&\hspace{1.5cm}\Bigl.-2s_1s_2 
\Bigl\{\Bigl(1+\frac{(x^1)^2+(x^2)^2}{z^2}\Bigr)\left(
\partial_\alpha x^1\partial_\beta x^1
+\partial_\alpha x^2\partial_\beta x^2
+\partial_\alpha z\partial_\beta z\right)\Bigr.\Bigr.\Bigr. \nonumber \\
&&\hspace{1.5cm}\Bigl.\Bigl.\Bigl.
-\Bigl(\partial_\alpha z+\frac{x^1}{z}\partial_\alpha x^1+\frac{x^2}{z}\partial_\alpha x^2\Bigr)
\Bigl(\partial_\beta z+\frac{x^1}{z}\partial_\beta x^1+\frac{x^2}{z}\partial_\beta x^2\Bigr)\Bigr\}\Bigr]\Bigr) \nonumber \\
&&+\int^\infty_{-\infty}\!\!\!d\tau\int^{2\pi}_0\!\!\!d\sigma~
\epsilon^{\alpha\beta}\frac{\eta}{z^4+2 s_1s_2 \eta^2 \left[z^2+(x^1)^2+(x^2)^2\right]} \nonumber \\
&&
\hspace{1cm}
\times\Bigl[x^1(s_1\partial_\alpha x^+ - s_2\partial_\alpha x^-)\partial_\beta x^1
+x^2(s_1\partial_\alpha x^+ - s_2\partial_\alpha x^-)\partial_\beta x^2\Bigr. 
\nonumber \\ &&\hspace{2cm}
\Bigl.+z(s_1\partial_\alpha x^+ - s_2\partial_\alpha x^-)\partial_\beta z\Bigr]\,. \label{action-3}
\end{eqnarray}
The action is complicated but it is still real.

\section{A solution in type IIB supergravity}

In this section we will present a solution in type IIB supergravity containing the metric 
and NS-NS two-form obtained from (\ref{action-1})\footnote{Note that the solutions with (\ref{action-2}) and (\ref{action-3}) 
will not be discussed hereafter,  because the metric is intricate and we have not succeeded to determine 
the other field components.}. 

\subsection{The action of type IIB supergravity}

Let us first introduce the equations of motion of type IIB supergravity \cite{IIB-SUGRA}. 
Here we will follow the notation of \cite{PS}.  
The action of the bosonic part is given by 
\begin{eqnarray}
S_{\rm IIB}=&&\frac{1}{2\kappa^2}\int\!d^{10}x\,\sqrt{-G}\,R
-\frac{1}{4\kappa^2}\int\Bigl(d\Phi\wedge*d\Phi
+{\rm e}^{2\Phi}dC\wedge*dC \Bigr. \nonumber \\
&&\Bigl.\qquad+{\rm e}^{-\Phi}H_3\wedge*H_3
+{\rm e}^{\Phi}\widetilde{F}_3\wedge*\widetilde{F}_3
+\frac{1}{2}\widetilde{F}_5\wedge*\widetilde{F}_5
+C_4\wedge H_3\wedge F_3\Bigr)\,. \label{SUGRA-action}
\end{eqnarray}
Here $G_{MN}$ is the 10D metric in the Einstein frame and  $R$ is its Ricci scalar. 
The constant parameter $\kappa$ is related to the 10D Newton constant $G_{10}$ like $2\kappa^2\equiv 16\pi G_{10}$\,.  
The symbol $\ast$ denotes the 10D Hodge dual operator.  
$\Phi$ is the fluctuation of the dilaton field and $C$ is the axion field. 
Then $B_2$\,, $C_2$ and $C_4$ are the NS-NS two-form, the R-R two-form 
and the R-R four-form. Their field strengths are defined as  
\begin{eqnarray}
H_3\equiv dB_2\,, \qquad F_3\equiv dC_2\,, \qquad F_5 \equiv dC_4
\end{eqnarray}
The modified field strengths $\widetilde{F}_3$ and $\widetilde{F}_5$ are defined as 
\begin{eqnarray}
&&\widetilde{F}_3 \equiv F_3-CH_3\,, \qquad 
\widetilde{F}_5 \equiv F_5-C_2\wedge H_3\,. 
\end{eqnarray}
Note that $\widetilde{F}_5$ has to satisfy the self-dual condition,
\begin{eqnarray}
\widetilde{F}_5=*\widetilde{F}_5\,. \label{self-dual}
\end{eqnarray}

\medskip 

By taking variations of the action (\ref{SUGRA-action})\,, the equations of motion 
are obtained as 
\begin{eqnarray}
&&R_{MN}=\frac{1}{2}\partial_M\Phi\partial_N\Phi+\frac{{\rm e}^{2\Phi}}{2}\partial_MC\partial_NC
+\frac{1}{96}\widetilde{F}_{MPQRS}\widetilde{F}_N^{~PQRS} \nonumber \\
&&\hspace{2.5cm}+\frac{1}{4}\left({\rm e}^{-\Phi}H_{MPQ}H_N^{~PQ}
+{\rm e}^{\Phi}\widetilde{F}_{MPQ}\widetilde{F}_N^{~PQ}\right) \nonumber \\
&&\hspace{2.5cm}-\frac{1}{48}G_{MN}\left({\rm e}^{-\Phi}H_{PQR}H^{PQR}
+{\rm e}^{\Phi}\widetilde{F}_{PQR}\widetilde{F}^{PQR}\right)\,, \label{eom:metric}\\ 
&&d*\widetilde{F}_5=-F_3\wedge H_3\,, \\ 
&&\nabla^2\Phi={\rm e}^{2\Phi}\partial^MC\partial_MC
-\frac{{\rm e}^{-\Phi}}{12}H_{MNP}H^{MNP}
+\frac{{\rm e}^{\Phi}}{12}\widetilde{F}_{MNP}\widetilde{F}^{MNP}\,, \\
&&\nabla^M({\rm e}^{2\Phi}\partial_MC)=-\frac{{\rm e}^{\Phi}}{6}H_{MNP}\widetilde{F}^{MNP}\,,  \\
&&d*({\rm e}^{-\Phi}H_3-{\rm e}^{\Phi}C\widetilde{F}_3)=-F_5\wedge F_3\,, \label{eom:B}\\ 
&&d*({\rm e}^{\Phi}\widetilde{F}_3)=F_5\wedge H_3\,. 
\end{eqnarray}
The Bianchi identities are given by 
\begin{eqnarray}
&&dH_3=0\,, \qquad dF_3=0\,, \qquad dF_5 =0\,, \\
&&d\widetilde{F}_3=-dC\wedge H_3\,, \qquad d\widetilde{F}_5=-F_3\wedge H_3\,. 
\end{eqnarray}
With this setup, we will consider a gravitational solution in the next subsection.

\subsection{A Jordanian deformed solution}

It is a turn to present a solution corresponding to a Jordanian deformation. 
From the construction of the string action, the metric is given by 
\begin{eqnarray}
ds^2&=&L^2\left[\frac{-2dx^+dx^-+(dx^1)^2+(dx^2)^2+dz^2}{z^2} \right. \\
&&\hspace{1cm} \left. -\frac{\eta^2}{z^4}\left(1+\frac{(x^1)^2+(x^2)^2}{z^2}\right)(dx^+)^2 
+ ds^2_{\rm S^5}\right]\,, \nonumber \\
ds^2_{\rm S^5}&=&ds^2_{{\mathbb C}\rm P^2}+(d\chi+\omega)^2\,. \nonumber
\end{eqnarray}
Here the metric of round S$^5$ is expressed as a $U(1)$ fibration over 
${\mathbb C}$P$^2$\,, where $\chi$ is the local coordinate on the Hopf fibre and $\omega$ 
is the one-form potential for the K\"ahler form on ${\mathbb C}$P$^2$\,. 
The metric of ${\mathbb C}$P$^2$ and $\omega$ are given by\footnote{We follow Appendix A.2 of \cite{ABM}.} 
\begin{eqnarray}
ds^2_{{\mathbb C}\rm P^2}&=&d\mu^2+\sin^2\mu\left(\Sigma_1^2+\Sigma_2^2+\cos^2\mu\,\Sigma_3^2\right)\,, \qquad
\omega=\sin^2\mu\,\Sigma_3\,, 
\end{eqnarray}
where $\Sigma_a~(a=1,2,3)$ are defined as 
\begin{eqnarray}
\Sigma_1& \equiv &\frac{1}{2}\left(\cos\psi\, d\theta +\sin\psi\sin\theta\, d\phi\right)\,, \quad 
\Sigma_2 \equiv \frac{1}{2}\left(\sin\psi\, d\theta -\cos\psi\sin\theta\, d\phi\right)\,, \nonumber \\
\Sigma_3& \equiv&\frac{1}{2}\left(d\psi+\cos\theta\, d\phi\right)\,. \nonumber
\end{eqnarray}
Note that the metric contains the 3D Schr\"odinger spacetime\footnote{
Therefore, the result of \cite{Kame} on the fast-moving limit \cite{Kruczenski} 
is directly applicable for this background. Note that the NS-NS two-form also vanishes at $x^1=x^2=0$\,. }
as a subspace with $x^1=x^2=0$\,, 
while the deformed AdS$_5$ part itself is not the 5D Schr\"odinger spacetime\footnote{
One may consider whether the deformed AdS$_5$ can be represented by a coset 
by following \cite{SYY}. }.  

\medskip 

Note that the metric which appear in the string action is represented in the string frame. 
However, the metric for the deformed AdS$_5$ part is invariant 
under the following scaling
\begin{eqnarray}
x^+ \to \lambda^2 x^+\,, \quad x^-\to x^-\,, \quad x^i \to \lambda x^i\,, 
\quad z \to \lambda z\,,
\end{eqnarray} 
and one may expect that the dilaton should be constant (i.e., $\Phi=0$). 
Thus the metric can be regarded as the one in the Einstein frame. 
For simplicity, we set $C=0$\,. 

\medskip 

By considering the S$^5$ components of the equation of motion for the metric (\ref{eom:metric}) 
and taking account of the self-duality condition (\ref{self-dual})\,, the five-form field-strength is fixed as 
\begin{eqnarray}
F_5=4L^4\left[ -\frac{1}{z^5}dx^+\wedge dx^-\wedge dx^1\wedge dx^2\wedge dz + {\rm vol}({\rm S^5})\right]\,. 
\end{eqnarray}
Thus $F_5$ is not modified under the deformation. 

\medskip 

Then the NS-NS two-form $B_2$ has also been derived as 
\begin{eqnarray}
B_2=\frac{L^2\eta}{z^4}\left(x^1dx^+\wedge dx^1+x^2dx^+\wedge dx^2\right)\,, 
\end{eqnarray}
and the associated field strength is given by 
\begin{eqnarray}
H_3=-\frac{4L^2\eta}{z^5}\left(x^1dx^+\wedge dx^1\wedge dz+x^2dx^+\wedge dx^2\wedge dz\right)\,. 
\end{eqnarray}
From the equation of motion for $H_3$\,, (\ref{eom:B})\,, 
one can notice that $F_3$ has to be turned on. 

\medskip 

The remaining task is to find out $F_3$ so as to satisfy all of the equations of motion. 
The resulting $F_3$ is give by 
\begin{eqnarray}
F_3 &=& \frac{4L^2\eta}{z^5}\left[x^2dx^+\wedge dx^1\wedge dz-x^1dx^+\wedge dx^2\wedge dz
-\frac{z}{2}dx^+\wedge dx^1\wedge dx^2\right] \nonumber \\ 
&& \hspace{1.5cm} - \frac{2L^2\eta}{z^3}\left[dx^+\wedge dz\wedge(d\chi+\omega)
-\frac{z}{2}dx^+\wedge d\omega\right]\,,  
\end{eqnarray}
where the associated R-R two-form $C_2$ is given by 
\begin{eqnarray}
C_2=-\frac{L^2\eta}{z^4}\left(x^2dx^+\wedge dx^1-x^1dx^+\wedge dx^2\right)
- \frac{L^2\eta}{z^2}dx^+\wedge(d\chi+\omega)\,. 
\end{eqnarray}

\medskip 

In summary, the gravitational solution in the Einstein frame is given by 
\begin{eqnarray}
ds^2&=&L^2\left[\frac{-2dx^+dx^-+(dx^1)^2+(dx^2)^2+dz^2}{z^2} \right. \nonumber \\
&&\hspace{1cm} \left. -\frac{\eta^2}{z^4}\left(1+\frac{(x^1)^2+(x^2)^2}{z^2}\right)(dx^+)^2 
+ ds^2_{{\mathbb C}{\rm P^2}}+(d\chi+\omega)^2\right]\,,  \label{sol} \\ 
F_5&=& 4L^4\left[-\frac{1}{z^5}dx^+\wedge dx^-\wedge dx^1\wedge dx^2\wedge dz + {\rm vol}({\rm S^5})\right]\,, 
\nonumber \\ 
H_3 &=& -\frac{4L^2\eta}{z^5}\left(x^1dx^+\wedge dx^1\wedge dz+x^2dx^+\wedge dx^2\wedge dz\right)\,, \nonumber \\ 
F_3 &=& \frac{4L^2\eta}{z^5}\left[x^2dx^+\wedge dx^1\wedge dz-x^1dx^+\wedge dx^2\wedge dz
-\frac{z}{2}dx^+\wedge dx^1\wedge dx^2\right] \nonumber \\ 
&& \hspace{1.5cm} - \frac{2L^2\eta}{z^3}\left[dx^+\wedge dz\wedge(d\chi+\omega)
-\frac{z}{2}dx^+\wedge d\omega\right]\,. \nonumber 
\end{eqnarray}
where  $\Phi=C=0$\,. The R-R scalar field $C$ may take a non-vanishing constant $C\neq 0$\,. 
In this case, the R-R two-form $C_2$ has to be shifted as $C_2 \to C_2 + C B_2$\,.  

\medskip 

Note that the Green-Schwarz string action on this solution (at the quadratic order of fermions) 
is easily obtained by substituting the solution (\ref{sol}) into (3.27) of \cite{GS-string-IIB}. 
Recently, the quartic-order action has been derived in \cite{fourth}. It would also be useful for further studies. 
As a matter of course, the total action is real, including the fermionic sector.  
It would be interesting to argue the world-sheet S-matrix by using the obtained action.

\paragraph*{The symmetry of the solution}  

Let us check the symmetry of the solution (\ref{sol}). 

\medskip 

We first concentrate on the symmetry of the deformed AdS part. 
It is obvious to see that the solution is invariant under 
two translations: 
\begin{eqnarray}
H:~x^+ \to x^+ + a^+\,, \qquad  M:~x^- \to x^- + a^-\,,
\end{eqnarray}
where $a^{\pm}$ are constant parameters. The invariance under the rotation in the 1-2 plane 
is also manifest. Recall that the solution is invariant under the anisotropic scaling, 
\begin{eqnarray}
D:~x^+ \to \lambda^2 x^+\,, \quad 
x^- \to x^-\,, \quad x^i \to \lambda x^i\,, \quad 
z\to \lambda z ~~~~~(\lambda:~\mbox{a constant})\,. 
\end{eqnarray}
A less obvious one is the special conformal transformation\footnote{For the derivation of this 
transformation law, for example, see \cite{Son}.},
\begin{eqnarray}
C: && x^+ \to (1-ax^+)x^+\,, \quad 
x^- \to x^- - \frac{a}{2}(x^ix^i+z^2)\,, 
\nonumber \\  
&& x^i \to (1-ax^+)x^i\,, \qquad z \to (1-ax^+)z\,,  
\end{eqnarray}
where $a$ is an infinitesimal parameter. 
Note that the solution (\ref{sol}) is not invariant under 
spatial translations and Galilean boosts due to the deformation. 
The symmetries $H$, $D$ and $C$ generate $SL(2,\mathbb{R})$\,. 
Then $M$ and the rotation in the 1-2 plane generates two $U(1)$'s. 

\medskip 

For the sphere part, the $SO(6)$ symmetry is broken to $SU(3) \times U(1)$ due to the 
presence of the R-R three-form field strength, where $SU(3)$ is the isometry of 
$\mathbb{C}$P$^2$ and $U(1)$ corresponds to a shift symmetry of $\chi$\,. 

\medskip 

In total, the resulting symmetry is given by 
\begin{eqnarray}
\left[SL(2,\mathbb{R})\times U(1)^2\right] \times \left[SU(3)\times U(1)\right]\,. 
\end{eqnarray}
It seems likely that the solution (\ref{sol}) is not supersymmetric 
because the $F_3$ flux is the same type of the one considered 
in \cite{MMT}, where the $H_3$ flux is considered but the mechanism 
to break supersymmetries would be identical. It might be interesting 
to consider a brane-wave deformation, instead of the $F_3$ flux, as in \cite{HY}. 
Some of the original supersymmetries may be preserved, 
while the integrability would become unclear. 

\medskip 

In comparison to the Jordanian deformed solution (\ref{sol}), it seems quite difficult to find out the full 
gravitational solution corresponding to the standard deformation in type IIB supergravity. 
The metric in the string frame is obtained in \cite{ABF}, but it involves a curvature singularity and 
the dilaton would be very complicated. 

\subsection{The tidal force}

It is also important to check whether the solution (\ref{sol}) involves a singularity or not. 
The solution is just regarded as a pp-wave like deformation of the AdS$_5\times$S$^5$ background. 
Hence no obvious curvature singularity is not found by computing curvature invariants. 
However, there may be another kind of singularity called pp-singularity \cite{pp}. 
In order to discuss this singularity, it is necessary to check the tidal force.  

\medskip 

First of all, one needs to take a time-like world-line and its tangent vector is 
\begin{eqnarray}
t^m
&=&\dot{x}^+\left(\frac{\partial}{\partial x^+}\right)^m
+\dot{x}^-\left(\frac{\partial}{\partial x^-}\right)^m
+\dot{x}^1\left(\frac{\partial}{\partial x^1}\right)^m
+\dot{x}^2\left(\frac{\partial}{\partial x^2}\right)^m
+\dot{z}\left(\frac{\partial}{\partial z}\right)^m\,, \nonumber 
\end{eqnarray}
where the index $m$ runs only for the deformed AdS$_5$ part and 
``dot'' denotes the derivative with respect to the affine parameter $\lambda$\,. 
Assume that the affine parameter is chosen so that the tangent vector becomes a unit vector: 
\begin{eqnarray}
G_{mn}t^m t^n=-1\,. 
\label{normalization1}
\end{eqnarray}

\medskip 

The dynamics of a particle moving on the solution (\ref{sol}) is described by the action
\begin{eqnarray}
S=\frac{1}{2}\int d\lambda~\frac{1}{z^2}\left[-2\dot{x}^+\dot{x}^-+(\dot{x}^1)^2+(\dot{x}^2)^2+\dot{z}^2
-\frac{\eta^2}{z^2}\left(1+\frac{(x^1)^2+(x^2)^2}{z^2}\right)(\dot{x}^+)^2\right]\,. \nonumber 
\end{eqnarray}
The equations of motion for $x^\pm$ provide two constants of motion, $P_-$ and $E$\,, 
\begin{eqnarray}
P_-=-\frac{\dot{x}^+}{z^2}\,, \qquad
E=\frac{1}{z^2}\left[-\dot{x}^- -\frac{\eta^2}{z^2}\left(1+\frac{(x^1)^2+(x^2)^2}{z^2}\right)\dot{x}^+\right]\,. 
\end{eqnarray}
Solving $P_-$ and $E$ with respect to $\dot{x}^\pm$ leads to the following expressions, 
\begin{eqnarray}
\dot{x}^+ =-z^2P_-\,, \qquad
\dot{x}^- =-z^2E+\eta^2\left(1+\frac{(x^1)^2+(x^2)^2}{z^2}\right)P_-\,. \label{pm}
\end{eqnarray}
The equations of motion for $x^1$ and $x^2$ are given by 
\begin{eqnarray}
\frac{d}{d\lambda}\left(\frac{\dot{x}^1}{z^2}\right)=-\frac{\eta^2x^1}{z^6}(\dot{x}^+)^2\,, \qquad
\frac{d}{d\lambda}\left(\frac{\dot{x}^2}{z^2}\right)=-\frac{\eta^2x^2}{z^6}(\dot{x}^+)^2\,. 
\end{eqnarray}
Noting that the normalization condition \eqref{normalization1} is explicitly written as 
\begin{eqnarray}
\frac{1}{z^2}\left[-2\dot{x}^+ \dot{x}^- +(\dot{x}^1)^2+(\dot{x}^2)^2+\dot{z}^2
-\frac{\eta^2}{z^2}\left(1+\frac{(x^1)^2+(x^2)^2}{z^2}\right)(\dot{x}^+)^2\right]=-1\,, 
\end{eqnarray}
one can solve (\ref{normalization1}) for $\dot{z}$ and obtain the following expression,  
\begin{eqnarray}
\dot{z}=\sqrt{-z^2+2EP_-z^4-\eta^2\left(z^2+(x^1)^2+(x^2)^2\right)P_-^2-(\dot{x}^1)^2-(\dot{x}^2)^2}\,. 
\end{eqnarray}
Here we have used the expressions of $\dot{x}^{\pm}$ given in (\ref{pm})\,. 

\medskip 

To evaluate the tidal force, it is not necessary to solve equations of motion explicitly.  
The tidal force is represented by the components of Riemann tensor 
in an orthonormal frame which is parallelly transported along the world-line.  
Thus one just needs to identify a basis $e^m$ for the orthonormal frame 
\begin{eqnarray}
\frac{d}{d\lambda}e^m = \Gamma^m_{np}\, t^n e^p\,.
\end{eqnarray} 
The orthonormal system is given by 
\begin{eqnarray}
n_1^m &=&
-\frac{\dot{x}^1}{P_-z}\left(\frac{\partial}{\partial x^-}\right)^m
+z\left(\frac{\partial}{\partial x^1}\right)^m\,, \qquad \nonumber \\ 
n_2^m &=&
-\frac{\dot{x}^2}{P_-z}\left(\frac{\partial}{\partial x^-}\right)^m
+z\left(\frac{\partial}{\partial x^2}\right)^m\,, \nonumber \\
p^m
&=&-\sin\lambda\Bigl[\dot{x}^+ \left(\frac{\partial}{\partial x^+}\right)^m
+\left(\dot{x}^- +\frac{1}{P_-}\right)\left(\frac{\partial}{\partial x^-}\right)^m
+\dot{x}^1\left(\frac{\partial}{\partial x^1}\right)^m
+\dot{x}^2\left(\frac{\partial}{\partial x^2}\right)^m \nonumber \\
&&\hspace{2cm}+\dot{z}\left(\frac{\partial}{\partial z}\right)^m 
\Bigr]
+\cos\lambda\Bigl[\frac{\dot{x}^-}{P_-z}\left(\frac{\partial}{\partial x^-}\right)^m
-z\left(\frac{\partial}{\partial z}\right)^m \Bigr]\,, \nonumber \\
q^m
&=&\cos\lambda\Bigl[\dot{x}^+\left(\frac{\partial}{\partial x^+}\right)^m
+\left(\dot{x}^- +\frac{1}{P_-}\right)\left(\frac{\partial}{\partial x^-}\right)^m
+\dot{x}^1\left(\frac{\partial}{\partial x^1}\right)^m
+\dot{x}^2 \left(\frac{\partial}{\partial x^2}\right)^m \nonumber \\
&&\hspace{2cm}+\dot{z}\left(\frac{\partial}{\partial z}\right)^m
\Bigr]
+\sin\lambda\Bigl[\frac{\dot{x}^-}{P_-z}\left(\frac{\partial}{\partial x^-}\right)^m
-z\left(\frac{\partial}{\partial z}\right)^m\Bigr]\,. \nonumber 
\end{eqnarray}
Then the tidal force is defined as 
\begin{eqnarray}
R_{(t)(e_1)(t)(e_2)}\equiv R^m_{~pqr}\,G_{mn}\,t^n\, e_1^p\, t^q\, e_2^r\,,
\end{eqnarray}
and the components of the tidal force are listed below\,:
\begin{eqnarray}
&&R_{(t)(1)(t)(1)}=1+2\eta^2\left(1+\frac{(x^1)^2+(x^2)^2}{z^2}\right)P_-^2\,, \qquad
R_{(t)(1)(t)(2)}=0\,, \nonumber \\
&&R_{(t)(2)(t)(2)}=1+2\eta^2\left(1+\frac{(x^1)^2+(x^2)^2}{z^2}\right)P_-^2\,, \nonumber \\
&&R_{(t)(1)(t)(p)}=4\eta^2\frac{x^1}{z}P_-^2\cos\lambda\,, \qquad
R_{(t)(1)(t)(q)}=4\eta^2\frac{x^1}{z}P_-^2\sin\lambda\,, \nonumber \\
&&R_{(t)(2)(t)(p)}=4\eta^2\frac{x^2}{z}P_-^2\cos\lambda\,, \qquad
R_{(t)(2)(t)(q)}=4\eta^2\frac{x^2}{z}P_-^2\sin\lambda\,, \nonumber \\
&&R_{(t)(p)(t)(p)}=1+4\eta^2\left(1+3\frac{(x^1)^2+(x^2)^2}{z^2}\right)P_-^2\cos^2\lambda\,, \nonumber \\
&&R_{(t)(p)(t)(q)}=4\eta^2\left(1+3\frac{(x^1)^2+(x^2)^2}{z^2}\right)P_-^2\sin\lambda\cos\lambda\,, \nonumber \\
&&R_{(t)(q)(t)(q)}=1+4\eta^2\left(1+3\frac{(x^1)^2+(x^2)^2}{z^2}\right)P_-^2\sin^2\lambda\,. \nonumber 
\end{eqnarray}
From the tidal force, one can see that the solution (\ref{sol}) is regular at the horizon, $z=\infty$\,, 
while it hits on a singular at the boundary, $z=0$\,, except at $x^1=x^2=0$\,. 

\medskip 

It is worth noting the similarity to the 5D Schr\"odinger spacetime with the dynamical critical exponent $z_{\rm c}$\,.  
When $z_{\rm c}=2$\,, there is no divergence of the tidal force at the horizon and the boundary \cite{Blau}.    
But, when $z_{\rm c}=3$\,, the tidal force diverges at the boundary \cite{Blau}. 
The solution (\ref{sol}) exhibits an isotropic scaling with $z_{\rm c}=2$\,, while its asymptotic behavior 
around the boundary is close to the one with $z_{\rm c}=3$\,. The divergence of the tidal force 
at the boundary in the solution (\ref{sol}) is similar to the one of the Schr\"odinger spacetime with $z_{\rm c}=3$\,.

\section{Conclusion and discussion}

We have considered a Jordanian deformation of the AdS$_5\times$S$^5$ superstring 
action with a simple $R$ operator satisfying CYBE. 
The metric and NS-NS two-form have explicitly been derived with a coordinate 
system. Only the AdS$_5$ part is deformed and the resulting geometry 
contains the 3D Schr\"odinger spacetime as a subspace.  
Then we have presented a solution in type IIB supergravity by determining the other field 
components. 
In particular, the dilaton is constant and a R-R three-form field strength is turned on. 
The symmetry of the solution is given by $\left[SL(2,\mathbb{R})\times U(1)^2\right]$ 
$\times$ $\left[SU(3)\times U(1)\right]$ and contains an anisotropic scale symmetry. 
Though the curvature invariants are not singular, 
the tidal force diverges at the boundary, except a certain point. 

\medskip 

There are many open problems now. The first is to consider a relation to deformed S-matrices on 
the string world-sheet. The standard $q$-deformations of the S-matrices are studied in 
\cite{BK,HHM,dLRT,Arutyunov}, but Jordanian deformed S-matrices have not been argued yet. 
It would be interesting to study them and compare the results with the string world-sheet S-matrices as in \cite{ABF}.  
The most important issue is the deformation of $\mathcal{N}$=4 SYM corresponding 
to the gravitational solution presented here. Probably, it would be concerned with non-local field theories such as 
dipole theories \cite{Ganor}. Although we have considered a deformation of the AdS$_5$ part, it might be possible to consider 
a similar deformation of the S$^5$ part. As far as we have tried, the metric contains imaginary parts 
and it seems difficult to give a physical interpretation. Anyway, because it should be regarded as a marginal deformation, 
such a complex solution might be related to 
a complex $\beta$-deformation discussed in \cite{LM}. 

\medskip 

The solution presented here is just an example. We expect that many interesting gravitational 
solutions would be found through Jordanian deformations. 
The recipe to look for them is given in \cite{KMY-Jordanian-typeIIB} and this paper. 
We hope that many integrable solutions are discovered

\subsection*{Acknowledgments}

The work of I.K.\ was supported by the Japan Society for the Promotion of Science (JSPS). 
T.M.\ also thanks G.~Arutyunov and R.~Borsato for useful discussions.  
T.M.\ is supported by the Netherlands Organization for Scientific 
Research (NWO) under the VICI grant 680-47-602.  
T.M.'s work is also part of the ERC Advanced grant research programme 
No.~246974, ``Supersymmetry: a window to non-perturbative physics" 
and of the D-ITP consortium, a program of the NWO that is funded by the 
Dutch Ministry of Education, Culture and Science (OCW).

\appendix

\section*{Appendix} 

\section{Our notation and convention}

Our notation and convention is summarized here by basically following \cite{AF-review}\,. 

\medskip 

An element of Lie superalgebra $\mathfrak{su}(2,2|4)$ is represented by an $8\times 8$ supermatrix: 
\begin{eqnarray}
M=\begin{bmatrix}
~m~&~\xi~\\
~\zeta~&~n~
\end{bmatrix}\,. 
\end{eqnarray}
Here $m$ and $n$ are $4\times 4$ matrices with Grassmann even elements, 
while $\xi$ and $\zeta$ are $4\times 4$ matrices with Grassmann odd elements. 
These matrices satisfy an appropriate reality condition. 
As a result, it turns out that $m$ and $n$ belong to $\mathfrak{su}(2,2)=\mathfrak{so}(2,4)$ and  
$\mathfrak{su}(4)=\mathfrak{so}(6)$\,, respectively.  

\medskip 

For our purpose, it is helpful to prepare an explicit basis of $\mathfrak{su}(4)$ and $\mathfrak{su}(2,2)$\,. 
Let us first introduce the following $\gamma$ matrices: 
\begin{eqnarray}
&&\gamma_1=
\begin{bmatrix}
~0~&~0~&~0~&-1~\\
~0~&~0~&~1~&~0~\\
~0~&~1~&~0~&~0~\\
-1~&~0~&~0~&~0~\\
\end{bmatrix}\,, \quad
\gamma_2=
\begin{bmatrix}
~0~&~0~&~0~&~i~\\
~0~&~0~&~i~&~0~\\
~0~&-i~&~0~&~0~\\
-i~&~0~&~0~&~0~\\
\end{bmatrix}\,, \quad 
\gamma_3=
\begin{bmatrix}
~0~&~0~&~1~&~0~\\
~0~&~0~&~0~&~1~\\
~1~&~0~&~0~&~0~\\
~0~&~1~&~0~&~0~\\
\end{bmatrix}\,, \nonumber \\
&&\gamma_4=
\begin{bmatrix}
~0~&~0~&-i~&~0~\\
~0~&~0~&~0~&~i~\\
~i~&~0~&~0~&~0~\\
~0~&-i~&~0~&~0~\\
\end{bmatrix}\,, \qquad
\gamma_5=-\gamma_1\gamma_2\gamma_3\gamma_4=
\begin{bmatrix}
~1~&~0~&~0~&~0~\\
~0~&~1~&~0~&~0~\\
~0~&~0~&-1~&~0~\\
~0~&~0~&~0~&-1~\\
\end{bmatrix}\,. 
\end{eqnarray}
Then $n_{ij}$ ($i,j=1,2,3,4,5$) are given by 
\begin{eqnarray}
n_{ij}=\frac{1}{4}\left[\gamma_i,\gamma_j\right]\,. 
\end{eqnarray}
It is easy to see that $\gamma_i$'s generate the Clifford algebra of $\mathfrak{so}(5)$: 
\begin{eqnarray}
\left\{\gamma_i,\gamma_j\right\}=2\delta_{ij}\,. 
\end{eqnarray}
Thus $n_{ij}$'s generate the Lie algebra $\mathfrak{so}(5)$\,. 
Note that
 \[
n_{ij}\,, \qquad n_{i6}=\frac{i}{2}\gamma_i
\] 
are regarded as the generators of $\mathfrak{so}(6)$\,. 

\medskip 

On the other hand, $\gamma_1$\,, $\gamma_2$\,, $\gamma_3$\,, $\gamma_0=i\gamma_4$ and $\gamma_5$ 
\begin{eqnarray}
&&\gamma_1=
\begin{bmatrix}
~0~&~0~&~0~&-1~\\
~0~&~0~&~1~&~0~\\
~0~&~1~&~0~&~0~\\
-1~&~0~&~0~&~0~\\
\end{bmatrix}\,, \quad
\gamma_2=
\begin{bmatrix}
~0~&~0~&~0~&~i~\\
~0~&~0~&~i~&~0~\\
~0~&-i~&~0~&~0~\\
-i~&~0~&~0~&~0~\\
\end{bmatrix}\,, \quad 
\gamma_3=
\begin{bmatrix}
~0~&~0~&~1~&~0~\\
~0~&~0~&~0~&~1~\\
~1~&~0~&~0~&~0~\\
~0~&~1~&~0~&~0~\\
\end{bmatrix}\,, \nonumber \\
&&\gamma_0=
\begin{bmatrix}
~0~&~0~&~1~&~0~\\
~0~&~0~&~0~&-1~\\
-1~&~0~&~0~&~0~\\
~0~&~1~&~0~&~0~\\
\end{bmatrix}\,, \qquad
\gamma_5=i\gamma_1\gamma_2\gamma_3\gamma_0=
\begin{bmatrix}
~1~&~0~&~0~&~0~\\
~0~&~1~&~0~&~0~\\
~0~&~0~&-1~&~0~\\
~0~&~0~&~0~&-1~\\
\end{bmatrix}\,. 
\end{eqnarray}
generate the Clifford algebra of $\mathfrak{so}(1,4)$\,: 
\begin{eqnarray}
&&\left\{\gamma_\mu,\gamma_\nu\right\}=2\eta_{\mu\nu} \qquad (\,\mu,\nu=0,1,2,3\,)\,, \\
&&\left\{\gamma_\mu,\gamma_5\right\}=0\,, \qquad 
\left(\gamma_5\right)^2=1\,. \nonumber 
\end{eqnarray}
Then the generators 
\begin{eqnarray}
n_{\mu\nu}=\frac{1}{4}\left[\gamma_\mu,\gamma_\nu\right]\,, \qquad 
n_{\mu 5}=\frac{1}{4}\left[\gamma_\mu,\gamma_5\right] 
\end{eqnarray}
satisfy the defining relations of $\mathfrak{so}(1,4)$\,. 
In addition, 
 \[
n_{\mu\nu}\,, \quad n_{\mu 5}\,, \quad
\gamma_\mu\,, \quad \gamma_5 
\]
are regarded as the spinor representation of $\mathfrak{so}(2,4)$\,.

\section{A list of $r$-matrices and deformed string actions} 

This appendix gives a list of lsome possible $r$-matrices and the associated string actions.

\medskip 

The AdS part of the Jordanian deformed action can be rewritten as 
\begin{align}
S_{\rm AdS}&=-\frac{1}{2}\int\!d\sigma^2(\ga^{\al\be}-\ep^{\al\be})
\Tr\left(A_\al P_2\circ  \frac{1}{1-2\eta [R_{\rm Jor}]_g\circ  P_2}A_\be\right) \nonumber \\
&=-\frac{1}{2}\int\!d\sigma^2(\ga^{\al\be}-\ep^{\al\be})
\Tr\left( A_\al P_2(J_\be ) \right) \nonumber \\
&=-\frac{1}{2}\int\!d\sigma^2(-\Tr( A_t P_2(J_t ) )+\Tr( A_x P_2(J_x ))) \el\\
&\quad +\frac{1}{2}\int\!d\sigma^2(\Tr( A_t P_2(J_x ) )-\Tr( A_x P_2(J_t))) \nonumber \\
&=\int\!d\sigma^2(L_{G}+L_{B})\,,   
\end{align}
where the sigma model part $L_G$ and the coupling to NS-NS two-form $L_B$ are given by  
\begin{align}
L_{G}&\equiv\frac{1}{2}\bigl[\Tr( A_t P_2(J_t ) )-\Tr( A_x P_2(J_x ))\bigr]\,, 
\el \\
L_{B} &\equiv \frac{1}{2}\bigl[\Tr( A_t P_2(J_x ) )-\Tr( A_x P_2(J_t ))\bigr]\,. 
\end{align}
The undeformed AdS$_5$ part is represented by 
\begin{align}
L_{G}^{\eta=0}&=-\frac{\ga^{\al\be}}{2z^2}
\left(-2\partial_\al x^+\partial_\be x^-
+\partial_\al x^1\partial_\be x^1
+\partial_\al x^2\partial_\be x^2 +\partial_\al z\partial_\be z\right)\,.  
\end{align}
This part is common for all of the deformation, and $L_{B}$ always vanishes in the $\eta\to0$ limit. 

\medskip 

It would be interesting to classify possible $r$-matrices and the associated string actions, 
though the classification here is forcussed upon some simple examples and not complete.  
Remarkably, all of the string actions contained in the list are real,  
up to surface terms appearing in $L_B$\,, after performing appropriate Wick rotations.  

\medskip 

The deformed string actions are classified into the three classes:
\begin{enumerate}
\item \qquad Class A$~=~\bigl\{(0), ~(1),~ (2),~ (3),~ (4)\bigr\}$\,,
\item \qquad Class B$~=~\bigl\{(5), ~(6)\bigr\}$\,, 
\item \qquad Class C$~=~\bigl\{(7),~ (8) \bigr\}$\,.  
\end{enumerate}
Each of the classes has the identical action, up to double Wick rotations and coordinate transformations. 
The class A corresponds to the case of \eqref{r-1} discussed in the body. 
The class B is the one discussed in subsection 2.3. 
The class C seems unphysical because two time directions appear  
after performing double Wick rotations to make the actions real. 

\medskip 

The three classes are listed below. 
\paragraph{Class A}  
\begin{enumerate}
\item[(0)] \qquad $r^{(0)}_{\rm Jor}=\displaystyle{\frac{1}{\sqrt{2}}}E_{24}\wedge(E_{22}-E_{44})$ \vspace*{5mm}\\ 
The deformed Lagrangian:
\begin{align}
L_{G}&=L_{G}^{\eta=0}+\eta^2 \ga^{\al\be} \frac{(x^1)^2+(x^2)^2+z^2}{2z^6}
\partial_\al x^+\partial_\be x^+\,,  \el \\
L_{B}&=\ep^{\al\be}\frac{\eta}{z^4}\partial_\al x^+
\left(x^1\partial_\be x^1+x^2\partial_\be x^2+z\partial_\be z\right)\,. \label{(0)}
\end{align}
This is the case with \eqref{r-1} considered in the body. The last term 
in $L_B$ is a surface term. It can be ignored without boundaries. 
The Lagrangian (\ref{(0)}) is invariant under $SL(2,\mathbb{R})\times U(1)^2$\,, 
which contains the anisotropic scaling invariance under 
\begin{eqnarray}
x^{+} \to \lambda^2 x^+\,, \quad x^- \to x^-\,, \quad 
x^i \to \lambda x^i\,, \quad z \to \lambda z\,, \label{scale}
\end{eqnarray}
where $\lambda$ is a constant. For the detail, see subsection 3.2. 

\item[(1)] \qquad  $r^{(1)}_{\rm Jor}=\displaystyle{\frac{1}{\sqrt{2}}}E_{13}\wedge(E_{11}-E_{33}) $ \vspace*{5mm}\\ 
The deformed  Lagrangian:
\begin{align}
L_{G}&=L_{G}^{\eta=0}+\eta^2 \ga^{\al\be} \frac{(x^1)^2+(x^2)^2+z^2}{2z^6}
\partial_\al x^-\partial_\be x^-\,,  \el \\
L_{B}&=-\ep^{\al\be}\frac{\eta}{z^4}\partial_\al x^-
\left(x^1\partial_\be x^1+x^2\partial_\be x^2+z\partial_\be z\right)\,. 
\end{align}
This can be obtained from the case (0) by exchanging  $x^{\pm} \to x^{\mp} $ 
and flipping $\eta \to -\eta$\,. Thus this case is equivalent to the case (0).

\item[(2)] \qquad $r^{(2)}_{\rm Jor}=\displaystyle{\frac{1}{\sqrt{2}}}E_{23}\wedge(E_{22}-E_{33}) $ \vspace*{0.5cm}\\ 
The deformed  Lagrangian:
\begin{align}
L_{G}&=L_{G}^{\eta=0}+\eta^2 \ga^{\al\be} \frac{-2x^+x^- +z^2}{2z^6}
\partial_\al \left(\frac{x^1-i x^2}{\sqrt{2}}\right)\partial_\be \left(\frac{x^1-i x^2}{\sqrt{2}}\right)\,,  \el \\
L_{B}&= -\ep^{\al\be}\frac{\eta}{z^4}\partial_\al \left(\frac{x^1-i x^2}{\sqrt{2}}\right)
(x^+\partial_\be x^-+x^-\partial_\be x^+-z\partial_\be z)\,. 
\end{align}
Note that $x^+\partial_\be x^-+x^-\partial_\be x^+= x^0\partial_{\beta}x^0 -x^3 \partial_{\beta}x^3$\,. 
After performing the double Wick rotation $x^2 \to ix^2$ and $x^0 \to ix^0$ 
and redefining the light-cone coordinates like $\tilde{x}^{\pm} = (x^2 \pm x^1)/\sqrt{2}$\,, 
this case is identical to the case (0), up to the total derivative.

\item[(3)] \qquad $r^{(4)}_{\rm Jor}=\displaystyle{\frac{1}{\sqrt{2}}}E_{14}\wedge(E_{11}-E_{44})$ \vspace*{0.5cm}\\ 
The deformed Lagrangian:
\begin{align}
L_{G}&=L_{G}^{\eta=0}+\eta^2 \ga^{\al\be} \frac{-2x^+x^- +z^2 }{2z^6}
\partial_\al \left(\frac{x^1+i x^2}{\sqrt{2}}\right)\partial_\be \left(\frac{x^1+i x^2}{\sqrt{2}}\right)\,,  \el \\
L_{B}&=\ep^{\al\be}\frac{\eta}{z^4}\partial_\al \left(\frac{x^1+i x^2}{\sqrt{2}}\right)
\left(x^+\partial_\al x^-+x^-\partial_\al x^+-z\partial_\al z\right)\,. \label{(3)}
\end{align}
After flipping $x^2 \to -x^2$ and $\eta \to -\eta$\,, this case is equivalent to the case (2). 
Thus the Lagrangian (\ref{(3)}) is also equivalent to the case (0), up to the total derivative.

\item[(4)] \qquad $r^{(3)}_{\rm Jor}=\displaystyle{\frac{1}{\sqrt{2}}}
\Bigl[E_{14}\wedge(E_{11}-E_{44})-2(E_{12}\wedge E_{24}+E_{13}\wedge E_{34}) \Bigr]$ 
\vspace*{0.5cm}\\ 
The deformed Lagrangian:
\begin{align}
L_{G} &=L_{G}^{\eta=0}+\eta^2 \ga^{\al\be} \frac{-2x^+x^- +z^2}{2z^6}
\partial_\al \left(\frac{x^1+i x^2}{\sqrt{2}}\right)\partial_\be \left(\frac{x^1+i x^2}{\sqrt{2}}\right)\,,  
\el  \\
L_{B} &=-\ep^{\al\be}\frac{\eta}{z^4}\partial_\al \left(\frac{x^1+i x^2}{\sqrt{2}}\right)
(x^+\partial_\be x^-+x^-\partial_\be x^++z\partial_\be z)\,. \label{(4)}
\end{align}
After flipping $x^2 \to -x^2$\,, this case is equivalent to the case (2), up to the total derivative. 
Thus the Lagrangian (\ref{(4)}) is also equivalent to the case (0)\,. 
\end{enumerate}
Note that the actions in the class A are identical, up to the total derivative. 
If boundaries are taken into account, the class A should be divided into subclasses. 
But we are interested in closed strings here and will not argue such subclasses.

\paragraph{Class B}
\begin{enumerate}  

\item[(5)] \qquad $r^{(5)}_{\rm Jor}=\displaystyle{\frac{i}{\sqrt{2}}}
\Bigl[E_{24}\wedge(E_{22}-E_{44})-2E_{23}\wedge E_{34}\Bigr] $ \vspace*{0.5cm}\\   
The deformed Lagrangian: 
\begin{align}
L_{\rm G}&=\frac{z^4}{z^4+4\eta^2(x^+)^2} 
\Bigl[L_{\rm G}^{\eta=0} 
-\frac{\eta^2}{2z^6}\gamma^{\alpha\beta}\Bigl(
4x^+\partial_\alpha x^+\left[x^1\partial_\beta x^1+x^2\partial_\beta x^2-2x^+\partial_\beta x^-\right]\bigr. \nonumber \\
&\hspace{4.3cm}\bigl.+4(x^+)^2\partial_\alpha z\partial_\beta z
+\left[z^2-(x^1)^2-(x^2)^2\right]\partial_\alpha x^+\partial_\beta x^+
\Bigr) \nonumber \\
&\hspace{3.5cm}\Bigl.
-2\frac{\eta^4}{z^8}(x^+)^2\gamma^{\alpha\beta}\partial_\alpha x^+\partial_\beta x^+\Bigr]\,,
\el \\
L_{\rm B}&=\frac{\eta}{z^4+4\eta^2(x^+)^2}\epsilon^{\alpha\beta} 
\Bigl[
x^2\partial_\alpha x^+\partial_\beta x^1
-x^1\partial_\alpha x^+\partial_\beta x^2
+2x^+\partial_\alpha x^1\partial_\beta x^2 \Bigr] \nonumber \\
&\quad+i\frac{\eta}{z^3}\epsilon^{\alpha\beta}\partial_\alpha x^+\partial_\beta z\,. \label{(5)}
\end{align}
The last term in $L_B$ is imaginary but just a surface term. 
Thus the Lagrangian (\ref{(5)}) is real without boundaries. 
Note that the Lagrangian (\ref{(5)}) is invariant under the anisotropic scaling (\ref{scale}), 
the rotation in the 1-2 plane and the shift of $x^-$\,, i.e., $U(1)^3$\,.

\item[(6)] \qquad $r^{(6)}_{\rm Jor}=\displaystyle{\frac{i}{\sqrt{2}}}
\Bigl[E_{13}\wedge(E_{11}-E_{33})-2E_{12}\wedge E_{23}\Bigr] $ \vspace*{0.5cm}\\ 
The deformed  Lagrangian:
\begin{align}
L_{\rm G}&=\frac{z^4}{z^4+4\eta^2(x^-)^2} 
\Bigl[L_{\rm G}^{\eta=0} 
-\frac{\eta^2}{2z^6}\gamma^{\alpha\beta}\Bigl(
4x^-\partial_\alpha x^-\left[x^1\partial_\beta x^1+x^2\partial_\beta x^2-2x^+\partial_\beta x^-\right]\bigr. \nonumber \\
&\hspace{4.3cm}\bigl.+4(x^-)^2\partial_\alpha z\partial_\beta z
+\left[z^2-(x^1)^2-(x^2)^2\right]\partial_\alpha x^-\partial_\beta x^-
\Bigr) \nonumber \\
&\hspace{3.5cm}\Bigl.
-2\frac{\eta^4}{z^8}(x^-)^2\gamma^{\alpha\beta}\partial_\alpha x^-\partial_\beta x^-\Bigr]\,,
\el \\
L_{\rm B}&=\frac{-\eta}{z^4+4\eta^2(x^-)^2}\epsilon^{\alpha\beta} 
\Bigl[
x^2\partial_\alpha x^-\partial_\beta x^1
-x^1\partial_\alpha x^-\partial_\beta x^2
+2x^-\partial_\alpha x^1\partial_\beta x^2 \Bigr] \nonumber \\
&\quad-i\frac{\eta}{z^3}\epsilon^{\alpha\beta}\partial_\alpha x^-\partial_\beta z\,. 
\end{align}
Through exchanging $x^{\pm} \to x^{\mp}$ and flipping $\eta \to -\eta$\,, this is equivalent to the case (5). 
\end{enumerate}
The class B corresponds to the case discussed in subsection 2.3.

\paragraph{Class C}
\begin{enumerate}  

\item[(7)] \qquad  $r_{\rm Jor}^{(7)}=\displaystyle{\frac{1}{\sqrt{2}}}
E_{34}\wedge(E_{33}-E_{44})$ \vspace*{0.5cm}\\ 
The deformed  Lagrangian: 
\begin{align}
L_{G}&=L_{G}^{\eta=0}-\eta^2 \ga^{\al\be} \frac{(x^1)^2+(x^2)^2-2x^+x^- }{2z^6}
\bigl[x^+\partial_\al (x^1+i x^2)-(x^1+i x^2)\partial_\al x^+ \bigl] \el \\
&\hspace{65mm}\times \bigl[x^+\partial_\be (x^1+i x^2)-(x^1+i x^2)\partial_\be x^+ \bigr]\,,  
\el \\
L_{B}&=\ep^{\al\be}\frac{\eta}{z^4}
\Bigl[
(x^1+ix^2)x^+\left(\partial_\alpha x^-\partial_\beta x^+
+i\partial_\alpha x^1\partial_\beta x^2\right) \nonumber \\
&\qquad\qquad 
+(x^1+ix^2)\partial_\alpha x^+\left(ix^2\partial_\beta x^1-ix^1\partial_\beta x^2\right) \nonumber \\
&\qquad\qquad 
-x^+\left(x^+\partial_\alpha x^- 
-x^-\partial_\alpha x^+\right)\partial_\beta(x^1+ix^2) 
\Bigr]\,. \label{(7)}
\end{align}
By performing a Wick rotation $x^2 \to -ix^2$\,, the Lagrangian (\ref{(7)}) becomes real 
but contain two time directions. Thus it seems to be unphysical. 
Note that, in comparison to the other cases, the Lagrangian (\ref{(7)}) is invariant under the isotropic scaling 
\[
x^{+} \to \lambda x^+\,, \quad x^- \to \lambda x^-\,, \quad 
x^i \to \lambda x^i\,, \quad z \to \lambda z\,,
\]
where $\lambda$ is a constant. After the Wick rotation, the Lagrangian (\ref{(7)}) is invariant 
also under the transformation,
\begin{eqnarray}
x^{+} \to \lambda'{} x^{+}\,, \quad x^{-} \to \lambda'{}^{-1} x^{-}\,, \quad 
\tilde{x}^{+} \to \lambda'{}^{-1} \tilde{x}^{+}\,, \quad \tilde{x}^{-} \to \lambda'{} \tilde{x}^{-}\,, 
\quad z \to z\,, \nonumber 
\end{eqnarray}
where $\tilde{x}^{\pm} = (x^2 \pm x^1)/\sqrt{2}$ and $\lambda'$ is a constant. 
This can be understood as the diagonal part of the two Lorentz boosts. 
In addition, it has the invariance under a ``rotation'' 
in the $(x^+,\tilde{x}^+)$ and $(x^-,\tilde{x}^-)$ planes,  
\[
x^\pm \to \cos\theta\, x^\pm - \sin\theta\, \tilde{x}^\pm\,, \qquad 
\tilde{x}^\pm \to \sin\theta\, x^\pm + \cos\theta\, \tilde{x}^\pm\,. 
\]
Thus the resulting symmetry is $U(1)^3$\,.

\item[(8)] \qquad $r_{\rm Jor}^{(8)}=\displaystyle{\frac{1}{\sqrt{2}}}
E_{12}\wedge(E_{11}-E_{22})$ \vspace*{0.5cm}\\ 
The deformed  Lagrangian: 
\begin{align}
L_{G}&=L_{G}^{\eta=0}-\eta^2 \ga^{\al\be} \frac{(x^1)^2+(x^2)^2-2x^+x^- }{2z^6}
\bigl[x^-\partial_\al (x^1+i x^2)-(x^1+i x^2)\partial_\al x^-\bigr] \el \\
&\hspace{65mm}\times \bigl[x^-\partial_\be (x^1+i x^2)-(x^1+i x^2)\partial_\be x^-\bigr]\,, \el \\
L_{B}&=\ep^{\al\be}\frac{\eta}{z^4}
\Bigl[
(x^1+ix^2)x^-\left(\partial_\alpha x^+\partial_\beta x^-
+i\partial_\alpha x^1\partial_\beta x^2\right) \nonumber \\
&\qquad\qquad 
+(x^1+ix^2)\partial_\alpha x^-\left(ix^2\partial_\beta x^1-ix^1\partial_\beta x^2\right) \nonumber \\
&\qquad\qquad 
-x^-\left(x^-\partial_\alpha x^+ 
-x^+\partial_\alpha x^-\right)\partial_\beta(x^1+ix^2) 
\Bigr]\,. 
\end{align}
By exchanging $x^{\pm} \to x^{\mp}$\,, this case is equivalent to the case (7). 
\end{enumerate}
The class C seems to be unphysical because of two time directions. 
It would be interesting to figure out a general criterion for the physical metric 
so as to exclude the class C.


\begin{thebibliography}{99}

\bibitem{M}  
  J.~M.~Maldacena,
  ``The large N limit of superconformal field theories and supergravity,''
  Adv.\ Theor.\ Math.\ Phys.\  {\bf 2} (1998) 231
  [Int.\ J.\ Theor.\ Phys.\  {\bf 38} (1999) 1113]. 
  [arXiv:hep-th/9711200].

\bibitem{GKP}
S.~S.~Gubser, I.~R.~Klebanov and A.~M.~Polyakov,
``Gauge theory correlators from non-critical string theory,''
Phys.\ Lett.\ B {\bf 428} (1998) 105 [arXiv:hep-th/9802109]. 

\bibitem{W}
E.~Witten, 
``Anti-de Sitter space and holography,''
Adv.\ Theor.\ Math.\ Phys.\  {\bf 2} (1998) 253 [arXiv:hep-th/9802150].

\bibitem{review}
  N.~Beisert {\it et al.},
  ``Review of AdS/CFT Integrability: An Overview,'' 
  Lett.\ Math.\ Phys.\ {\bf 99} (2012) 3 [arXiv:1012.3982 [hep-th]]. 

\bibitem{MT}
  R.~R.~Metsaev and A.~A.~Tseytlin,
  ``Type IIB superstring action in AdS$_5\times$S$^5$ background,''  
  Nucl.\ Phys.\ B {\bf 533} (1998) 109  [hep-th/9805028].  

\bibitem{BPR}
  I.~Bena, J.~Polchinski and R.~Roiban,
  ``Hidden symmetries of the AdS$_5\times$S$^5$ superstring,''
  Phys.\ Rev.\ D {\bf 69} (2004) 046002
  [hep-th/0305116].
  
\bibitem{RS}  
 R.~Roiban and W.~Siegel,
  ``Superstrings on AdS$_5\times$S$^5$ supertwistor space,''  
JHEP {\bf 0011} (2000) 024  [hep-th/0010104].  
  
\bibitem{Hatsuda}
  M.~Hatsuda and K.~Yoshida,
  ``Classical integrability and super Yangian of superstring on AdS$_5\times$S$^5$,''  
Adv.\ Theor.\ Math.\ Phys.\  {\bf 9} (2005) 703  [hep-th/0407044];   
  ``Super Yangian of superstring on AdS$_5\times$S$^5$ revisited,''  
Adv.\ Theor.\ Math.\ Phys.\  {\bf 15} (2011) 1485  [arXiv:1107.4673 [hep-th]].

\bibitem{Zarembo-symmetric}
  K.~Zarembo,
  ``Strings on Semisymmetric Superspaces,''
  JHEP {\bf 1005} (2010) 002
  [arXiv:1003.0465 [hep-th]]. 

\bibitem{Wulf}
  L.~Wulff,
  ``Superisometries and integrability of superstrings,''  arXiv:1402.3122 [hep-th]. 
  
  
  
  
\bibitem{Drinfeld1}
  V.~G.~Drinfel'd,
  ``Hopf algebras and the quantum Yang-Baxter equation,'' 
  Sov.\ Math.\ Dokl.\ {\bf 32} (1985) 254.  

\bibitem{Drinfeld2}
 V.~G.~Drinfel'd,
  ``Quantum groups,''
  J.\ Sov.\ Math.\  {\bf 41} (1988) 898 
  [Zap.\ Nauchn.\ Semin.\  {\bf 155}, 18 (1986)].

\bibitem{Jimbo}
  M.~Jimbo,
  ``A $q$ difference analog of $U(g)$ and the Yang-Baxter equation,''
  Lett.\ Math.\ Phys.\  {\bf 10} (1985) 63.      

\bibitem{CP} 
 V.~Chari and A.~N.~Pressley,   
 ``A Guide to Quantum Groups,"    
 Cambridge University Press.     
  
  
\bibitem{Cherednik}
  I.~V.~Cherednik, 
  ``Relativistically Invariant Quasiclassical Limits Of Integrable
  Two-Dimensional Quantum Models,''
  Theor.\ Math.\ Phys.\  {\bf 47} (1981) 422
  [Teor.\ Mat.\ Fiz.\  {\bf 47} (1981) 225].

\bibitem{FR}
  L.~D.~Faddeev and N.~Y.~Reshetikhin,
  ``Integrability of the principal chiral field model in (1+1)-dimension,''
  Annals Phys.\  {\bf 167} (1986) 227.      

\bibitem{BFP}
  J.~Balog, P.~Forgacs and L.~Palla,
  ``A two-dimensional integrable axionic sigma model and T duality,''  
  Phys.\ Lett.\ B {\bf 484} (2000) 367  
  [hep-th/0004180].  

\bibitem{Klimcik}
 C.~Klimcik,
  ``Yang-Baxter sigma models and dS/AdS T duality,''  
JHEP {\bf 0212} (2002) 051  [hep-th/0210095]; 
  ``On integrability of the Yang-Baxter sigma-model,''  
J.\ Math.\ Phys.\  {\bf 50} (2009) 043508  [arXiv:0802.3518 [hep-th]].

\bibitem{Mohammedi}
  N.~Mohammedi,
  ``On the geometry of classically integrable two-dimensional non-linear sigma models,''
  Nucl.\ Phys.\ B {\bf 839} (2010) 420
  [arXiv:0806.0550 [hep-th]].  

\bibitem{KY}
  I.~Kawaguchi and K.~Yoshida,
  ``Hidden Yangian symmetry in sigma model on squashed sphere,''
  JHEP {\bf 1011} (2010) 032. 
  [arXiv:1008.0776 [hep-th]]. 

\bibitem{KYhybrid}
  I.~Kawaguchi and K.~Yoshida,
  ``Hybrid classical integrability in squashed sigma models,''
  Phys.\ Lett.\ B\ {\bf 705} (2011) 251
  [arXiv:1107.3662 [hep-th]]; 
   ``Hybrid classical integrable structure of squashed sigma models: A short summary,''  
  J.\ Phys.\ Conf.\ Ser.\  {\bf 343} (2012) 012055 
  [arXiv:1110.6748 [hep-th]].    
  
\bibitem{KMY-QAA}
  I.~Kawaguchi, T.~Matsumoto and K.~Yoshida,
  ``The classical origin of quantum affine algebra in squashed sigma models,''  
  JHEP {\bf 1204} (2012) 115  [arXiv:1201.3058 [hep-th]].  

\bibitem{KMY-monodromy}
  I.~Kawaguchi, T.~Matsumoto and K.~Yoshida,
  ``On the classical equivalence of monodromy matrices in squashed sigma model,''  
  JHEP {\bf 1206} (2012) 082  [arXiv:1203.3400 [hep-th]].

\bibitem{ORU}
  D.~Orlando, S.~Reffert and L.~I.~Uruchurtu,
  ``Classical integrability of the squashed three-sphere, warped AdS3 and
  Schr$\ddot{\rm o}$dinger spacetime via T-Duality,''
  J.\ Phys.\ A  {\bf 44} (2011) 115401.
  [arXiv:1011.1771 [hep-th]]. 
  
\bibitem{KOY}
  I.~Kawaguchi, D.~Orlando and K.~Yoshida,
  ``Yangian symmetry in deformed WZNW models on squashed spheres,''
  Phys.\ Lett.\  B {\bf 701} (2011) 475. 
  [arXiv:1104.0738 [hep-th]]; 
  I.~Kawaguchi and K.~Yoshida,
  ``A deformation of quantum affine algebra in squashed WZNW models,''
  arXiv:1311.4696 [hep-th].

\bibitem{BR}
   B.~Basso and A.~Rej,
  ``On the integrability of two-dimensional models with $U(1) \times SU(N)$ symmetry,''  
   Nucl.\ Phys.\ B {\bf 866} (2013) 337  [arXiv:1207.0413 [hep-th]]. 
 
\bibitem{DMV}
  F.~Delduc, M.~Magro and B.~Vicedo,
  ``On classical q-deformations of integrable sigma-models,''  
  JHEP {\bf 1311} (2013) 192  [arXiv:1308.3581 [hep-th]]. 

\bibitem{Squellari}
  R.~Squellari,
  ``Yang-Baxter $\sigma$ model: Quantum aspects,''  arXiv:1401.3197 [hep-th]. 

\bibitem{bi-YB}
  C.~Klimcik, 
  ``Integrability of the bi-Yang-Baxter sigma model,'' arXiv:1402.2105 [math-ph]. 
 
\bibitem{DMV2}
  F.~Delduc, M.~Magro and B.~Vicedo,
  ``An integrable deformation of the AdS$_5\times$S$^5$ superstring action,''  
 Phys.\ Rev.\ Lett.\  {\bf 112} (2014) 051601
  [arXiv:1309.5850 [hep-th]].
  
\bibitem{ABF}
    G.~Arutyunov, R.~Borsato and S.~Frolov,
  ``S-matrix for strings on $\eta$-deformed AdS$_5\times$S$^5$,''
  arXiv:1312.3542 [hep-th]. 
  
\bibitem{R}
  N.~Reshetikhin,
  ``Multiparameter quantum groups and twisted quasitriangular Hopf algebras,''  
  Lett.\ Math.\ Phys.\  {\bf 20} (1990) 331.  
  
\bibitem{Jordanian}
  A.~Stolin and P.~P.~Kulish, 
  ``New rational solutions of Yang-Baxter equation and deformed Yangians,''
  Czech.\ J.\ Phys.\ {\bf 47} (1997) 123 [arXiv:q-alg/9608011].  

\bibitem{KLM}
  P.~P.~Kulish, V.~D.~Lyakhovsky and A.~I.~Mudrov,
  ``Extended jordanian twists for Lie algebras,''  
  J.\ Math.\ Phys.\  {\bf 40} (1999) 4569  [math/9806014 [math.QA]].      


 
\bibitem{Tolstoy}
 V.~N.~Tolstoy, ``Chains of extended Jordanian twists for Lie superalgebras,'' \\
 math/0402433 [math.QA]. 
   
\bibitem{BLT}
A.~Borowiec, J.~Lukierski and V.~N.~Tolstoy,
``New twisted quantum deformations of D=4 super-Poincare algebra,'' 
arXiv:0803.4167 [hep-th].

\bibitem{ACS} 
N.~Aizawa, R.~Chakrabarti and J.~Segar, 
``Jordanian Quantum \\ Superalgebra U$_{h}$(osp(2$|$1)),''
Mod.\ Phys.\ Lett.\ {\bf A18} (2003) 885-903, math/0301022. 

\bibitem{ACCYZ}
B.~Abdesselam, A.~Chakrabarti, R.~Chakrabarti, A.~Yanallah and M.~B.~Zahaf, 
``On super-Jordanian ${\cal U}_{\sf h}(sl(N|1))$ algebra,'' 
J.\ Phys.\ A: Math.\ Gen.\ {\bf 39} (2006) 8307-8319, math/0511430.
   
  
\bibitem{KMY-Jordanian-typeIIB}
  I.~Kawaguchi, T.~Matsumoto and K.~Yoshida,
  ``Jordanian deformations of the AdS$_5\times$S$^5$ superstring,''
  arXiv:1401.4855 [hep-th].

\bibitem{KY-Sch}
  I.~Kawaguchi and K.~Yoshida,
  ``Classical integrability of Schr\"odinger sigma models and $q$-deformed Poincare symmetry,''  
JHEP {\bf 1111} (2011) 094  [arXiv:1109.0872 [hep-th]]; 
  ``Exotic symmetry and monodromy equivalence in Schr\"odinger sigma models,''  
JHEP {\bf 1302} (2013) 024  [arXiv:1209.4147 [hep-th]].

\bibitem{Jordanian-KMY}
  I.~Kawaguchi, T.~Matsumoto and K.~Yoshida,
  ``Schr\"odinger sigma models and Jordanian twists,''  
JHEP {\bf 1308} (2013) 013  [arXiv:1305.6556 [hep-th]].  




\bibitem{IIB-SUGRA}
  J.~H.~Schwarz,
  ``Covariant Field Equations of Chiral N=2 D=10 Supergravity,''  
Nucl.\ Phys.\ B {\bf 226} (1983) 269.

\bibitem{PS}
  J.~Polchinski and M.~J.~Strassler,
  ``The String dual of a confining four-dimensional gauge theory,''  
hep-th/0003136. 

\bibitem{ABM}
 A.~Adams, K.~Balasubramanian and J.~McGreevy,
  ``Hot Spacetimes for Cold Atoms,''
  JHEP {\bf 0811} (2008) 059
  [arXiv:0807.1111 [hep-th]].


\bibitem{Kame}
  T.~Kameyama and K.~Yoshida,
  ``String theories on warped AdS backgrounds and integrable deformations of spin chains,''  
  JHEP {\bf 1305} (2013) 146  [arXiv:1304.1286 [hep-th]]. 
  
\bibitem{Kruczenski}
 M.~Kruczenski,
  ``Spin chains and string theory,''
  Phys.\ Rev.\ Lett.\  {\bf 93} (2004) 161602
  [hep-th/0311203].  

\bibitem{SYY}
  S.~Schafer-Nameki, M.~Yamazaki and K.~Yoshida,
  ``Coset Construction for Duals of Non-relativistic CFTs,''
  JHEP {\bf 0905} (2009) 038
  [arXiv:0903.4245 [hep-th]].         
 
\bibitem{GS-string-IIB}
  M.~Cvetic, H.~Lu, C.~N.~Pope and K.~S.~Stelle,
  ``T duality in the Green-Schwarz formalism, and the massless / massive IIA duality map,''
  Nucl.\ Phys.\ B {\bf 573} (2000) 149
  [hep-th/9907202]. 
  
\bibitem{fourth}
  L.~Wulff,
  ``The type II superstring to order $\theta^4$,''  
JHEP {\bf 1307} (2013) 123  [arXiv:1304.6422 [hep-th]].  

\bibitem{Son} 
 D.~T.~Son,
  ``Toward an AdS/cold atoms correspondence: A Geometric realization of the Schrodinger symmetry,''
  Phys.\ Rev.\ D {\bf 78} (2008) 046003
  [arXiv:0804.3972 [hep-th]].
  
\bibitem{MMT}
  J.~Maldacena, D.~Martelli and Y.~Tachikawa,
  ``Comments on string theory backgrounds with non-relativistic conformal symmetry,''
  JHEP {\bf 0810} (2008) 072
  [arXiv:0807.1100 [hep-th]].

\bibitem{HY}  
 S.~A.~Hartnoll and K.~Yoshida,
  ``Families of IIB duals for nonrelativistic CFTs,''
  JHEP {\bf 0812} (2008) 071
  [arXiv:0810.0298 [hep-th]].
  

\bibitem{pp}
  D.~Brecher, A.~Chamblin and H.~S.~Reall,
  ``AdS / CFT in the infinite momentum frame,''
  Nucl.\ Phys.\ B {\bf 607} (2001) 155
  [hep-th/0012076].

\bibitem{Blau}
  M.~Blau, J.~Hartong and B.~Rollier,
  ``Geometry of Schrodinger Space-Times, Global Coordinates, and Harmonic Trapping,''
  JHEP {\bf 0907} (2009) 027
  [arXiv:0904.3304 [hep-th]].
  
\bibitem{BK}
  N.~Beisert and P.~Koroteev,
  ``Quantum Deformations of the One-Dimensional Hubbard Model,''  
J.\ Phys.\ A {\bf 41} (2008) 255204  [arXiv:0802.0777 [hep-th]]. \\ 
 N.~Beisert, W.~Galleas and T.~Matsumoto, 
``A Quantum Affine Algebra for the Deformed Hubbard Chain,'' 
J.\ Phys.\ A {\bf 45} (2012) 365206 [arXiv:1102.5700 [math-ph]].  

\bibitem{HHM}
 B.~Hoare, T.~J.~Hollowood and J.~L.~Miramontes,
  ``$q$-Deformation of the AdS$_5\times$S$^5$ Superstring S-matrix 
and its Relativistic Limit,''  
JHEP {\bf 1203} (2012) 015  [arXiv:1112.4485 [hep-th]]; 
  ``Bound States of the $q$-Deformed AdS$_5\times$S$^5$ Superstring S-matrix,''  
JHEP {\bf 1210} (2012) 076  [arXiv:1206.0010 [hep-th]]; 
 ``Restoring Unitarity in the $q$-Deformed World-Sheet S-Matrix,'' 
 JHEP {\bf 1310} (2013) 050 [arXiv:1303.1447 [hep-th]].

\bibitem{dLRT}
  M.~de Leeuw, V.~Regelskis and A.~Torrielli,
  ``The Quantum Affine Origin of the AdS/CFT Secret Symmetry,''  
J.\ Phys.\ A {\bf 45} (2012) 175202  [arXiv:1112.4989 [hep-th]].

\bibitem{Arutyunov}
  G.~Arutyunov, M.~de Leeuw and S.~J.~van Tongeren,
  ``The Quantum Deformed Mirror TBA I,''  
JHEP {\bf 1210} (2012) 090 [arXiv:1208.3478 [hep-th]];  
 ``The Quantum Deformed Mirror TBA II,''  
JHEP {\bf 1302} (2013) 012 [arXiv:1210.8185 [hep-th]].


\bibitem{Ganor}
  M.~Alishahiha and O.~J.~Ganor,
  ``Twisted backgrounds, PP waves and nonlocal field theories,''  
JHEP {\bf 0303} (2003) 006  [hep-th/0301080].  
  
  

\bibitem{LM}
 O.~Lunin and J.~M.~Maldacena,
  ``Deforming field theories with $U(1) \times U(1)$ global symmetry and their gravity duals,''  
JHEP {\bf 0505} (2005) 033  [hep-th/0502086].


\bibitem{AF-review}
  G.~Arutyunov and S.~Frolov,
  ``Foundations of the AdS$_5\times$S$^5$ Superstring. Part I,''
  J.\ Phys.\ A {\bf 42} (2009) 254003
  [arXiv:0901.4937 [hep-th]].    


\end{thebibliography}
\end{document}